\documentclass {aa} 

\usepackage{graphicx}
\begin{document}


\title{Stellar sources in the ISOGAL intermediate bulge fields\thanks{This is 
paper no. 14 in a refereed journal based on data from the ISOGAL project.}
\fnmsep \thanks{Based on observations with ISO, an ESA project with instruments
funded by ESA Member States (especially the PI countries: France, Germany, the 
Netherlands and the United Kingdom) and with the participation of ISAS and 
NASA.}
\fnmsep \thanks {Based on observations collected at the European Southern
Observatory, La Silla Chile.}}

\author{
D. K. Ojha\inst{1}
\and A. Omont\inst{2}
\and F. Schuller\inst{2}
\and G. Simon\inst{3}
\and S. Ganesh\inst{4}
\and M. Schultheis \inst{2}
}

\offprints{D. K. Ojha, \email{ojha@tifr.res.in}}

\institute{
Tata Institute of Fundamental Research, Homi Bhabha Road, Colaba, 
Mumbai - 400 005, India\\
\email{ojha@tifr.res.in}
\and Institut d'Astrophysique de Paris, CNRS, 98bis Bd Arago, F-75014 Paris, 
France\\
\email{omont@iap.fr}
\and GEPI, Observatoire de Paris, 61, Av. de l'Observatoire, F-75014, Paris, 
France
\and Physical Research Laboratory, Navarangpura, Ahmedabad 380009, India
}


\date{Received 03 June 2002 / Accepted 11 February 2003}


\abstract{
We present a study of ISOGAL sources in the ``intermediate'' galactic bulge
($|$$l$$|$ $<$ 2$^\circ$, $|$$b$$|$ $\sim$ 1$^\circ$--4$^\circ$), observed by 
ISOCAM at 7 and 15 $\mu m$. In combination with near-infrared (I, J, 
K$_{\rm s}$) 
data of DENIS survey, complemented by 2MASS data, we discuss the nature 
of the ISOGAL sources, their luminosities, the interstellar extinction and the 
mass-loss rates. A large fraction of the 1464 detected sources at 15 $\mu m$ 
are AGB stars above the RGB tip, a number of them show an excess in 
([7]-[15])$_{\rm 0}$ and \mbox{(K$_{\rm s}$-[15])$_{\rm 0}$} colours, 
characteristic of 
mass-loss. The latter, especially (K$_{\rm s}$-[15])$_{\rm 0}$, provide 
estimates of the 
mass-loss rates and show their distribution in the range 
\mbox{10$^{-8}$ to 10$^{-5}$ M$_{\rm \odot}$/yr}.  
}

\maketitle

\section {Introduction}

ISOGAL is a detailed mid-infrared imaging survey of the inner Galaxy, tracing 
the Galactic structure and stellar populations (Omont et al. 2003). It 
combines 7 and 15 $\mu m$ ISOCAM data with IJK$_{\rm s}$ DENIS data 
(Epchtein et al. 1994, 1997; Simon et al. in preparation). 
The main goals of ISOGAL survey 
are to trace the large scale inner disk and bulge structure using primarily 
red giants old stars; to study the corresponding stellar populations; to 
trace the age and mass-loss of AGB stars; to determine the number of (dusty) 
young stars and to map the star formation regions through diffuse ISM emission 
and extinction. Multicolour near- and mid-infrared data are essential to
analyse these features. 

The scientific results of ISOGAL are summarized in Omont et al. (2003). The 
analysis of the first ISOGAL field is detailed in P\'erault et al. (1996). The 
discussion of a small field in the inner bulge 
($\ell$~=~0$^{\circ}$, $b$~=~+1$^{\circ}$) (Omont et al. 1999) has confirmed
and characterized the remarkable capability of ISOGAL to detect the mass-losing
AGB stars. Glass et al. (1999) analysed the ISOGAL fields in Baade's Windows 
of low obscuration towards the bulge. Most of the detected objects towards 
Baade's Windows are late-type M stars, with a cut-off for those earlier than 
about M3--M4. In the same fields, Alard et al. (2001) have determined the 
general properties of a well defined sample of semi-regular variables by 
obtaining the MACHO lightcurves in V and R for a large fraction ($\sim$ 300) 
of the ISOGAL objects in Baade's Windows. Recently, Schultheis \& Glass (2001) and Glass \& Schultheis (2002) 
have further investigated the luminous M-type giants in these Baade's Windows 
fields, in the near-infrared mainly taken from the DENIS and 2MASS 
(Skrutskie et al. 1995) surveys.

In this paper, we report the study of nine ISOGAL fields in the intermediate 
galactic bulge (--1.5$^{\circ}$$<$$\ell$$<$+1.6$^{\circ}$; 
--3.8$^{\circ}$$\le$$b$$\le$--1.0$^{\circ}$, $b$ = +1.0$^{\circ}$ (Fig. 1); 
with a total 
area $\sim$ 0.29 deg$^2$) and more than 2300 detected sources. The 15 $\mu m$ 
and 7 $\mu m$ ISOGAL point sources have been combined primarily with 
DENIS IJK$_{\rm s}$ data 
in the ISOGAL--DENIS Point Source Catalogue which gives
magnitudes I, J, K$_{\rm s}$, [7], [15] (Schuller et al. 2003), -- and also 
with 
2MASS (J, H, K$_{\rm s}$) and MSX (mid--IR, Price et al. 2001) -- to determine 
their 
nature and the interstellar extinction. Analysis at five 
near- and mid- infrared wavelengths of the ISOGAL sources, together with the 
relatively low and constant extinction, shows that the majority of the sources 
are red giants with luminosities just above or close to the RGB tip and with a 
large proportion of detectable mass-loss. The various colour-colour and 
magnitude-colour diagrams are discussed in the paper, together with values of 
the luminosities and mass-loss rates.

The outline of the paper is as follows : In Sect. 2, we present the ISOGAL and 
DENIS observations, as well as complementary 2MASS (and MSX) data. Sect. 3 
describes the cross-identification of the mid-IR ISOGAL sources between 7 \& 
15 $\mu m$ bands and then with near-infrared sources. In Sect. 4, 
we discuss the determination of interstellar extinction in the line of sight 
of the bulge fields using the isochrone fitting. Sect. 5 presents the 
derivation 
of bolometric magnitude (M$_{\rm bol}$) and luminosity for each star in the 
bulge 
fields. In Sect. 6, we discuss the nature of ISOGAL sources. Sect. 7 is devoted to the 
derivation of mass-loss rates based on \mbox{(K$_{\rm s}$-[15])$_{\rm 0}$} 
colour. 
In Sect. 8, we 
consider a preliminary estimation of the total mass-loss rate in the 
intermediate bulge.

\begin{figure}
\centering
\resizebox{\hsize}{!}{\includegraphics{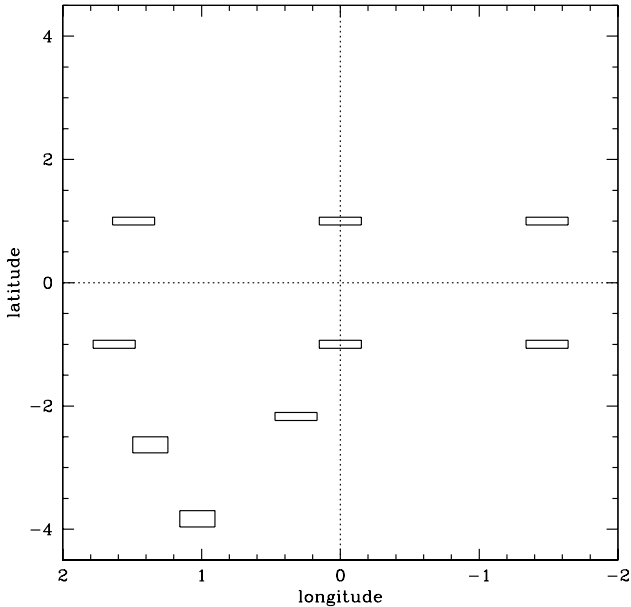}}
\caption{The rectangles show the limits of the ISOGAL fields
studied in this paper.}
\label{FIG1}
\end{figure}

\begin{figure*}
\centering
\resizebox{\hsize}{!}{\includegraphics{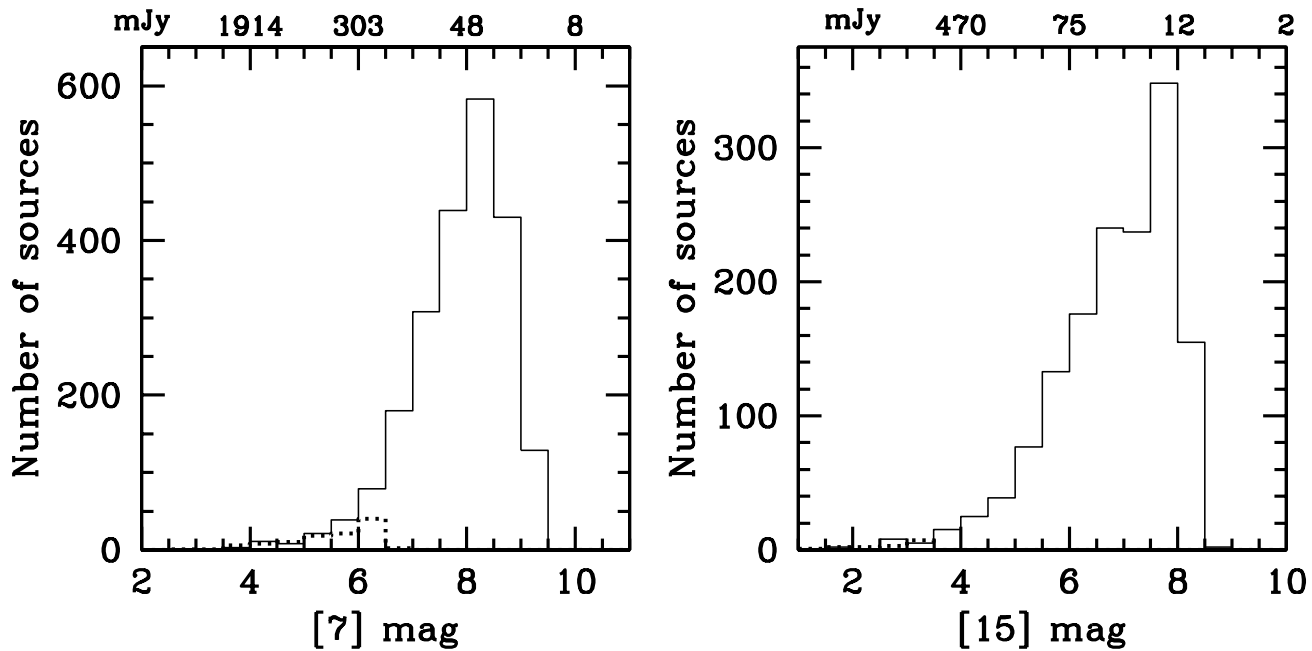}}
\caption{7 and 15 $\mu m$ source distributions in half magnitude bins. 
ISOGAL magnitudes are converted into flux densities using the relations given 
in Schuller et al. (2003). The dotted lines show the MSX A ($\sim$ 8 $\mu m$) 
and D ($\sim$ 15 $\mu m$) sources cross-identified with ISOGAL 7 and 
15 $\mu m$ sources. MSX magnitudes are derived using the zero magnitude 
flux given by Egan et al. (1999).}
\label{FIG2}
\end{figure*}

\section{ISOGAL observations and complementary data}

ISOGAL observations with ISOCAM (see C\'esarsky et al. 1996 for a general 
reference to ISOCAM operation and performances) at 6$^{''}$ pixel field of 
view of nine intermediate bulge fields with good quality of 7 and 15 $\mu m$ 
data are used in our study (see Fig. 1). These observations were performed 
during 1996-1998 at 15~$\mu m$ (filter LW3 : 12-18~$\mu m$ or 
LW9 : 14-16~$\mu m$), and at 7~$\mu m$ (filter LW2 : 5.5-8.5~$\mu m$ or 
LW6 : 7.0-8.5~$\mu m$) (Table 1). For all the observed ISOGAL fields, we have 
used the data of the ISOGAL Point Source Catalogue (PSC, Version 1) described 
in Schuller et al. (2003), reproduced, for our nine fields, in a shortened 
version in Table 2 available at CDS\footnote{Table 2 contains the catalogue of
ISOGAL-DENIS sources from nine bulge fields with their 2MASS \& MSX 
counterparts. The catalogue will be available via the VizieR Service at the 
Centre de Donn\'ees Astronomiques de Strasbourg 
(CDS, http://vizier.u-strasbg.fr/viz-bin/VizieR).}. They were derived with an 
improved data 
reduction with respect to ISO Archive data, with the use of CIA ISOCAM 
software and of a special source extraction (Schuller et al. 2003). From the 
comparison between the different observations of the field at 
$l = 0^\circ$, $b = +1^\circ$, we find very small differences between the 
different filters : LW2 - LW6 = +0.05 mag \& \mbox{LW3 - LW9 = +0.1 mag}. 
Hence, no corrections are applied in LW6 and LW9 magnitudes to jointly discuss 
them with the main LW2 and LW3 data. The total number of sources in the nine 
fields are 2232 and 1464 at 7 and 15 $\mu m$, respectively, with a grand total of 2354 sources.
The corresponding average source densities are 
\mbox{$\sim$ 8$\times$10$^3$ deg$^{-2}$} and $\sim$ 5$\times$10$^3$ deg$^{-2}$ at 7 
and 15 $\mu m$, respectively (see detailed density for each field as shown in 
Table 1). The 7 and 15 $\mu m$ magnitude limits of the ISOGAL PSC are given in 
Table 1 and vary from 8.6 to 9.5 at 7 $\mu m$, and from 
\mbox{7.5 to 8.5 at 15 $\mu m$}. Generally, these ISO cutoff magnitudes should 
roughly correspond to a detection completeness level close to $\sim$ 50\%, and 
depend on the source density, so that these magnitude limits vary by about 
1 mag among the different fields. The histograms of the 7 and 15~$\mu m$ total 
source counts derived from the $6\arcsec$ ISOGAL observations are displayed in 
Fig. 2. This figure also displays the data (available at IPAC, version 
1.2) of the MSX sources detected in these fields 
(108 at 8~$\mu m$, all with [7]$_{\rm ISO}$ $<$ 7.0 and 16 at 
15~$\mu m$, all with [15])$_{\rm ISO}$ $<$ 4.5). They show the large difference 
in sensitivity and in the number of 
detected sources between the two surveys. We have checked the consistency of 
ISOGAL and MSX 15~$\mu m$ magnitudes for the 13 strongest sources 
(with [15]~$<$~3.5, in order to avoid the Malmquist bias). The average 
difference is [15]$_{\rm ISO}$--[15]$_{\rm MSX}$~=~--0.06$\pm$0.43, where 
the large 
dispersion is probably related to the source variability.

In the near-infrared, the  data  mostly used in this paper are also those of the PSC of the ISOGAL--DENIS survey. They are complemented by the 2MASS data (from the Point Source Catalog of the 2MASS second release 
available at IPAC and CDS). However, these 2MASS data do not fully cover 
all the nine ISOGAL fields. There are large blank areas
(because of saturating stars) in three fields, so that only about 88\% of the 
total area of the nine ISOGAL fields is covered. All sources in these 
2MASS data have detections in the three JHK$_{\rm s}$ bands. 
The DENIS near-infrared data are acquired from the
DENIS survey with special observations of the Galactic bulge (Simon et al.,
in preparation) in the three bands, I (0.80 $\mu m$), J (1.25 $\mu m$) and 
K$_{\rm s}$ (2.17 $\mu m$), which completely cover the nine considered ISOGAL 
bulge fields. Therefore, we have taken the more complete DENIS data as the 
main reference for near-infrared associations.

The histograms of all the DENIS I, J, K$_{\rm s}$ sources in the observed 
fields are shown in Fig. 3. The 
completeness limit is close to 12.0 in K$_{\rm s}$ band, 14.0 in J band and 
15.5 in 
I band, respectively. However, only near-infrared  sources with 
K$_{\rm s}$ $\la$ 11.0 are 
used for near-infrared cross-identifications, in order to limit the number of 
spurious associations (see Sect. 3 and Table 1). Fig. 3 also displays the 
histograms of J, H and K$_{\rm s}$ 2MASS sources. We have compared the DENIS
and 2MASS photometry of 1963 stars common in ISOGAL--DENIS PSC and 2MASS in our
nine fields. We find small differences in the averages: 
\mbox{K$_{\rm DENIS}$--K$_{\rm 2MASS}$ = -0.00$\pm$0.07} and 
J$_{\rm DENIS}$--J$_{\rm 2MASS}$ = -0.02$\pm$0.09; hence no corrections 
are applied. 

Table 1 shows the used ISOGAL, DENIS and 2MASS observations in detail. 
It should be
noted that the majority of the observations were performed on dates with 
large differences (from two months to three years), except for 7 and 15 
$\mu m$ observations performed on the same date in six fields out of nine, 
and the DENIS, 2MASS observations performed on the same date for all 
sources in their respective bands. 
Such differences in dates are comparable to the period of Long Period 
Variables (LPVs). One thus expects that the corresponding colours of the few 
large amplitud \& variables (Miras) may be strongly affected. This uncertainty 
will propagate through to the final uncertainties in luminosity and 
mass-loss as well. From the detailed studies of LPVs in the two Baade's Window 
fields (see Glass et al. 1999, Alard et al. 2001, Schultheis \& Glass 2001, 
Glass \& Schultheis 2002), one can estimate that the variability r.m.s. error 
in the near-infrared is $\simeq$~0.3 mag for strong variables (miras), i.e. 
most of the strong sources with [15]~$<$~4, and less than 0.1~mag for most 
others. In order to reduce this uncertainty by about a factor $\sqrt{2}$, we 
have systematically used: 1) The average of DENIS and 2MASS J and K$_{\rm s}$ 
magnitudes when available; however, the DENIS sources are 
saturated at K$_{\rm s}$ $\la$ 6.5, and, for such few strong sources, 
only 2MASS 
data are used (Sect. 3). 2) The average of ISOGAL and MSX 15 $\mu m$ magnitudes 
for the few (13) strong sources detected by MSX with [15]$_{\rm ISO}$~$<$~3.5. 

\begin{table*}
\caption[]{Log of ISOCAM, DENIS and 2MASS observations in the 
intermediate bulge fields}
\hspace{-1cm}
\vspace{0cm}
\begin{tabular}{l c c c c c c c c c}\\
\hline
ISOGAL & ISOGAL & Filter & Date of & Source       & 7 and      & Date of & Date of & Mean  & DENIS \\
Field  & Obs.   &        & ISOGAL  & Densities    & 15 $\mu m$ & DENIS   & 2MASS   & A$_{\rm V}$ & K$_{\rm s}{\rm _Lim}$\\
       &        &        & Obs.    & at 7 \& 15 $\mu m$ & mag  & IJK$_{\rm s}$ & JHK$_{\rm s}$    \\
       &        &        &         &(per deg$^2$)  & limits    & Obs. & Obs. \\
\hline
-01.49-01.00 & 83801111 & LW2 & 02/03/98 & 8500 & 9.03 &22/05/99 \& & 14/08/98 & 11.8 & 11.4 \\
             & 32500342 & LW3 & 06/10/96 & 5231 & 8.12 & 08/07/99\\
+00.00-01.00 & 84100926 & LW2 & 05/03/98 & 10607 & 8.80 & 01/05/99 & 16/07/98 & 8.1  & 10.5 \\
             & 84100927 & LW3 & 05/03/98 & 6929 & 7.93\\
+01.63-01.00 & 84101405 & LW6 & 05/03/98 & 6538 & 8.61 & 03/07/98 & 16/07/98 & 6.3  & 10.9 \\
             & 84101406 & LW9 & 05/03/98 & 3654 & 7.53 \\ 
-01.49+01.00 & 83701309 & LW2 & 01/03/98 & 9111 & 8.94 & 29/08/98 & 02/07/98 & 6.0  & 10.8 \\
             & 32500238 & LW3 & 06/10/96 & 6556 & 8.04 \\
+00.00+01.00 & 83600418 & LW2 & 28/02/98 & 9259 & 8.89 & 18/05/98 & 02/07/98 & 6.3  & 10.6 \\
             & 83600523 & LW3 & 28/02/98 & 6148 & 8.00 \\
+01.49+01.00 & 84001007 & LW6 & 04/03/98 & 7962 & 8.65 & 09/05/99 & 02/07/98 & 6.4  & 10.9 \\
             & 32500239 & LW3 & 06/10/96 & 6077 & 7.99 \\
+01.37-02.63 & 83800913 & LW2 & 02/03/98 & 6660 & 9.15 & 13/09/98 \& & 16/07/98 & 1.2 & 11.4 \\ 
(Sgr I BW field) & 83800914 & LW3 & 02/03/98 & 4453 & 8.33 & 20/06/99\\
+01.03-03.83 & 84001115 & LW2 & 04/03/98 & 5608 & 9.48 & 15/09/98 & 15/08/98 & 1.5  & 12.0 \\
(NGC 7522 BW & 84001116 & LW3 & 04/03/98 & 3333 & 8.51 \\
field) & \\
+00.32-02.17 & 84100428 & LW2 & 05/03/98 & 7519 & 9.14 & 20/05/98 & 16/07/98 & 2.2  &  11.0 \\
             & 84100429 & LW3 & 05/03/98 & 4889 & 8.26 \\
\hline
\label{table1}
\end{tabular}
\end{table*}

\begin{figure}
\centering
\resizebox{\hsize}{!}{\includegraphics{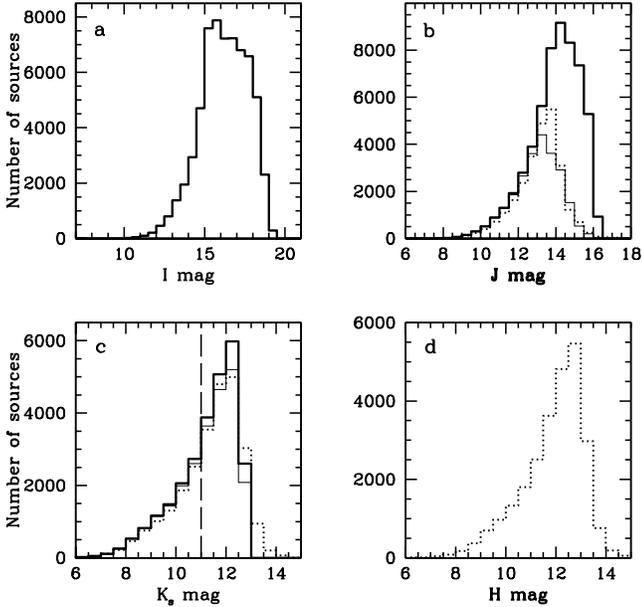}}
\caption{DENIS source counts in I, J \& K$_{\rm s}$ bands in half magnitude bins
(thick continuous lines). Thin continuous lines show the DENIS sources
detected in J and K$_{\rm s}$ bands. Dotted lines show the 2MASS sources
detected in J, H \& K$_{\rm s}$ bands. The used 2MASS data (2nd release)
contain only sources detected in the three JHK$_{\rm s}$ bands and cover
only $\sim$ 88\% of the total area of the considered ISOGAL fields.
Long-dashed line shows the average
K$_{\rm s}$ magnitude cut of DENIS sources  (K$_{\rm s}$ $<$ 11, see Table 1 and
Sect. 3) cross-identified with ISOGAL sources.}
\label{FIG3}
\end{figure}

\section{Cross-identification of ISOGAL and near-infrared sources}

In order to perform the association between 7 and 15 $\mu m$
sources in the ISOGAL PSC, first a global offset between 
the two sets of coordinates has been determined, then distortion coefficients 
have been computed 
to correct for the rotation effects (Schuller et al. 2003).
Finally, 7 and 15 $\mu m$ associations were searched within a radius of twice the size of the 
pixel (2$\times$6\arcsec = 12\arcsec). Only $\sim$ 8\% (122) of the 15 $\mu m$ 
sources have no 7 $\mu m$ counterpart. However, out of these 122 
15 $\mu m$ sources, a large fraction (77\% : 94 sources) have been associated 
with K$_{\rm s}$ (K$_{\rm s}$ $\la$ 11.0) DENIS or 2MASS sources, 
which reflects the incompleteness
of weak 7 $\mu m$ sources. The r.m.s. of the [7]-[15] separations  has a 
small value equal to 1.5\arcsec, which shows the good quality of most of the 
associations. Indeed, 98\% have a quality factor 3 or 4 in the ISOGAL PSC which
warrants almost certainly real association (Schuller et al. 2003), 
only 2\% have a quality factor 2 which means that the association is
probably real, but should be considered with some care (a few doubtful 
associations with quality factor 1 have been discarded). 

\begin{figure}
\centering
\resizebox{\hsize}{!}{\includegraphics{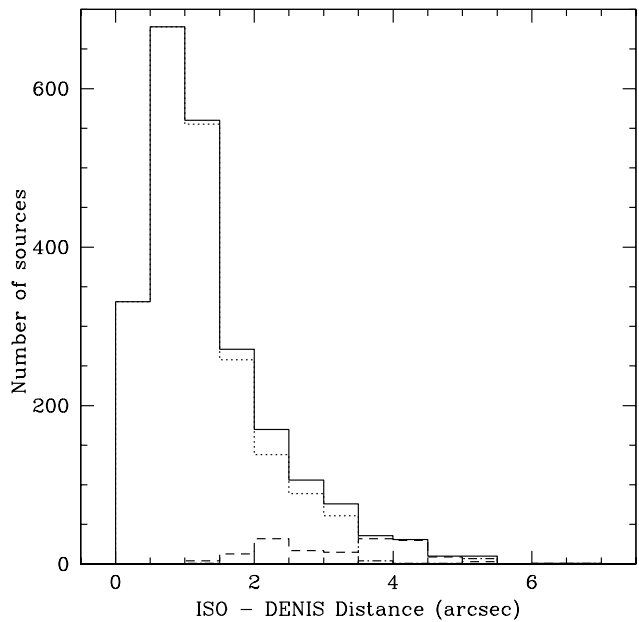}}
\caption{Histogram of positional difference (in arcsec) between ISOGAL and 
DENIS cross-identified sources. Solid line shows all ISOGAL/DENIS sources.
Dotted line shows the sources having ISOGAL/DENIS association quality
flags 4 \& 5. Dashed line shows the sources having quality flags 2 \& 3. 
Dotted-dashed line shows the sources having quality flag 1. Associations with quality flag 1 and those with a distance larger than 5\arcsec  have been finally discarded.} 
\label{FIG4}
\end{figure}

A cross-correlation algorithm similar to the one used to associate 7 and
15 $\mu m$ data has been used to look for DENIS counterparts to ISOGAL
sources (Schuller et al. 2003). The density of DENIS sources is so high that 
the DENIS catalogues were cut to 36000 sources per square degree 
(K$_{\rm s}$ $\la$ 11.0, see Table 1). A search radius of 7\arcsec ~was initially used for
the cross-correlation between ISOGAL and DENIS sources, 
and the association was limited to DENIS sources detected in K$_{\rm s}$.
Figure 4 shows the histogram of positional
difference of ISOGAL--DENIS cross-identified sources in our nine fields. The r.m.s. of the distances of ISOGAL-DENIS association has a  
value equal to 1.4\arcsec. About 98\% of the ISOGAL sources (either 7 or 15 $\mu m$) have thus initially an association with a DENIS 
source with 
K$_{\rm s}$~$\la$~11.0.
Even with such a large search radius, because the proportion of real associations is very large and only associations with the smallest separation are retained, the fraction of false
associations is expected to remain $\sim$1\%. 
In addition, the quality of the associations is characterised by
a specific flag and we have dropped in Table 2 the associations with the poorest quality (see below). Such a large radius was indeed
needed since, in some cases, real
associations have such large separation, because of the large pixels and undersampling of ISOGAL, in particular in
the cases of blended or slightly extended sources. This is extensively 
discussed in Schuller et al. (2003).  
About 1\% of the ISOGAL sources (23) have a DENIS 
counterpart which are saturated (K$_{\rm s}$ $\la$ 6.5, see Fig. 7),
out of which 18 saturated DENIS sources have been replaced by 
the 2MASS sources (as indicated in Table 2 available at CDS). The five other sources
are not identified in the published 2MASS fields. Out of these 18 sources,
50\% sources are identified as foreground and taken out of the stars further
discussed in the paper (see text, Sect. 4) and the
remaining are shown in various figures (see Figs. 8, 9, 11 \& 12). 
Most of the ISOGAL sources
cross-identified with DENIS ones (92\%) have a good quality association 
(quality flags 4 \& 5; see Schuller et al. 2003 for details), 7\% sources 
have more doubtful association (quality flags 2 \& 3) and only less than 1\% ones have 
very bad association (quality flag 1). For the present study, we have dropped in Table 2 the
sources which have association quality flag of 1 (15 associations)
and those with associated distance of more than 5\arcsec ~(12 associations, 
mostly with quality flag 2). The number of contaminating false matches is thus kept well below 1\%. 
The respective numbers of 7 $\mu m$ alone, 15 $\mu m$ alone \& 7+15 $\mu m$ sources not 
associated at all with DENIS sources are 21, 32 \& 18. However, most of these 7+15 $\mu m$ sources are not in the DENIS catalogue because they are too strong and above the saturation limit; 13 of them are present in the 2MASS data. Most 
(18) of the sources of 7~$\mu m$ alone are also associated with 2MASS sources 
with K$_{\rm s}$~$\la$~11.0, so that the proportion of 7~$\mu m$ sources not 
associated with a near-IR source is negligible ($\sim$ 0.4\%). On the other 
hand, only four sources of 15~$\mu m$ alone are associated with 2MASS sources 
with K$_{\rm s}$~$\la$~11.0; it results in a more important, but still small, 
fraction ($\sim$ 2\%) of  15 $\mu m$ sources without near-IR associations, 
especially for the sources detected only at 15 $\mu m$ ($\sim$ 23\%). 
The latter could be young stellar objects (YSOs, Sect.
 6.2); however, half of them have poor quality flags which could be 
characteristic of the spatial extension of YSOs (Schuller et al. in 
preparation) or cast a doubt on the reality of unassociated sources. Finally, 
taking into account the 15+12 rejected associations, the percentage of 
ISOGAL sources with a near-infrared association is larger than 97\%.

\begin{figure}
\centering
\resizebox{\hsize}{!}{\includegraphics{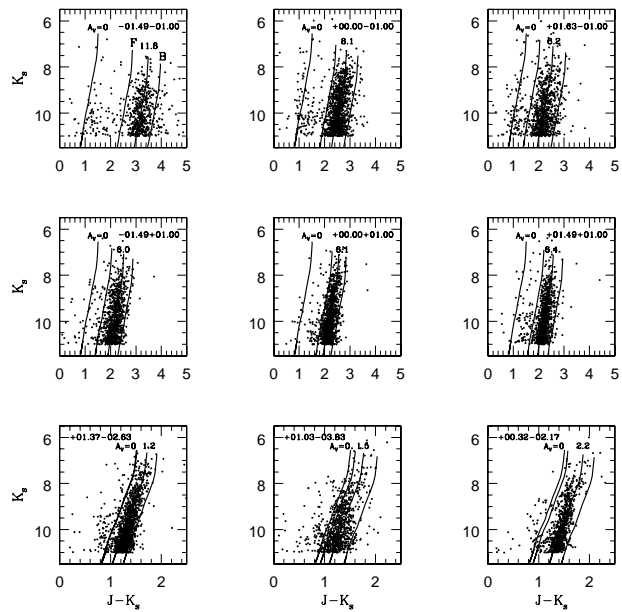}}
\caption{Colour-magnitude diagrams (J-K$_{\rm s}$/K$_{\rm s}$) of DENIS sources in the
ISOGAL intermediate bulge fields. The four isochrones (Bertelli et al. 1994), 
placed at 8 kpc distance for a 10 Gyr population with Z=0.02 are shown
in each figure for A$_{\rm V}$ = 0 mag, for the A$_{\rm V}$ limit adopted for the 
``foreground'' sources (defined as ``F'' in top left box), for mean A$_{\rm V}$ 
values of the field and for the A$_{\rm V}$ limit adopted for the ``background'' 
sources (defined as ``B''), respectively.} 
\label{FIG5}
\end{figure}

\section{Interstellar extinction and foreground disk stars}

\subsection{Interstellar extinction}

The adjunction of DENIS and 2MASS near-infrared data adds much to ISOGAL data, 
by providing different and more sensitive colour indices, as well as estimates
of the interstellar reddening. The K$_{\rm s}$/J-K$_{\rm s}$ magnitude-colour 
diagrams of DENIS sources in the bulge fields show a well-defined red giant 
sequence shifted by fairly uniform extinction (see Fig. 5),
with respect to the reference K$_{\rm s0}$ {\it vs} (J-K$_{\rm s}$)$_{\rm 0}$ 
of Bertelli et al (1994) with Z=0.02 and a distance modulus of 14.5 
(distance to the Galactic Center : 8 kpc). The near-infrared colours of this
isochrone have been computed with an empirical T$_{\rm eff}$-(J-K)$_{\rm 0}$ 
colour relation built by making a fit through measurements for cool giants 
(see Schultheis et al. 1998; 1999). We have assumed that 
A$_{\rm J}$/A$_{\rm V}$ = 0.256; A$_{\rm K}$$_{\rm s}$/A$_{\rm V}$ = 0.089 
(Glass 1999). The average value of the interstellar extinction
(A$_{\rm V}$) for each bulge field is shown in Table 1 and in Fig. 5. 
The mean A$_{\rm V}$
value for each field has been determined by a gaussian fit to the 
A$_{\rm V}$ distribution.

The DENIS sources with anomalously low values of A$_{\rm V}$ are
probably foreground. They are visible in Fig. 5 in each of the bulge fields 
(stars around the isochrone
with A$_{\rm V}$ $\sim$ 0 mag and stars clearly left of the bulk of the stars 
grouped around the isochrone with mean A$_{\rm V}$ of the field). 
For each field we define an isochrone ``F'' (Fig. 5) for which we assume
that all the sources left of it are foreground stars which will be no
longer considered in the following discussions of ``bulge'' stars.

There are also a number of stars with J-K$_{\rm s}$
values significantly larger than the bulk of the other stars in each bulge
field (right of the isochrone defined as ``B'' in Fig. 5). We can see three 
reasons for such an excess in J-K$_{\rm s}$ :
1) intrinsic (J-K$_{\rm s}$)$_{\rm 0}$ excess induced by a large mass-loss, 
which should
be accompanied by a large 15 $\mu m$ excess; 2) excess in A$_{\rm V}$ which 
should probably be due to a patchy extinction on the bulge line of
sight (additional extinction from dust layers behind the Galactic
Center for background stars appears unlikely at such high galactic
latitudes); 3) spurious association or wrong photometry which is
rather unlikely for DENIS sources well above the detection limit.

For the stars with such large values of J-K$_{\rm s}$, right of the
isochrone ``B'' in Fig. 5, we distinguish: a) bright stars, with 
K$_{\rm s}$ $<$ 8,
which can be bulge AGB stars with large mass-loss, for which we again
use the mean extinction correction A$_{\rm V}$ of the field; and b) faint
stars, with K$_{\rm s}$ $>$ 8, for which we assume low mass-loss and large 
extinction and we determine their specific extinction
from J-K$_{\rm s}$ and the zero-extinction isochrone as for the bulk
of bulge sources.

To summarize, for the following discussions where dereddening is essential -- 
colour--magnitude diagrams, luminosities, mass--loss rates, we have applied 
the following prescriptions : We have discarded all sources left of curves 
``F'' in Fig. 5. For the others, with J and K detections, besides the few 
exceptions mentioned above, we have determined their specific extinction 
from J-K$_{\rm s}$ [with Glass (1999) values for A$_{\rm J}$/A$_{\rm V}$ and 
A$_{\rm K}$$_{\rm s}$/A$_{\rm V}$] and the quoted zero-extinction isochrone of 
M giants. For all the others, especially with no JK associations, we have used 
the mean extinction A$_{\rm V}$ of the field (Table 1). The extinction ratios 
A$_{\rm [7]}$/A$_{\rm V}$ and A$_{\rm [15]}$/A$_{\rm V}$ in the ISOGAL bands 
are still 
uncertain (Hennebelle et al. 2001, Jiang et al. 2003). We have used the values 
A$_{\rm [7]}$/A$_{\rm V}$ = 0.020; A$_{\rm [15]}$/A$_{\rm V}$ = 0.025 
recommended by 
Hennebelle et al. for ``disk'' fields; however, the extinction corrections 
remain small with the low values of A$_{\rm V}$ considered, and the effects of 
the uncertainty on A$_{\rm [7]}$ and A$_{\rm [15]}$ are practically negligible.

\subsection{Proportion of foreground and background disk stars}

It is clear that the method discussed above for identifying foreground
stars from the value of J-K$_{\rm s}$ and the inferred extinction, fails to
recognize foreground disk stars located between the last dust layers
and the bulge. Similarly, the extinction gives no way of distinguishing
background disk stars, beyond the bulge, from bulge stars.

In order to derive an order of magnitude of the proportion of such
foreground and background disk stars, mixed with bulge stars with
similar extinction, one can use any reasonable simple Galactic model
for the M--K giants of the bulge and the disk, e.g. the one by Wainscoat
et al. (1992). In Appendix A, we have thus estimated the approximate fraction 
of counts expected from the bulge, the  disk r~$<$~R$_{\rm 0}$, and the  
disk r~$>$~R$_{\rm 0}$, as being about 65\%, 23\% and 12\%, respectively, on 
a typical line
of sight for our fields, $l = 0$, $b = 2^\circ$ (R$_0$ is the distance to 
the Galactic Center).

Indeed, stars of the actual bulge and of the central disk are
indistinguishable in our analysis. We should thus consider as really 
foreground and background, only disk stars outside of the bulge radius 
\mbox{$R_1$ = 2 kpc} of Wainscoat et al (1992). Their fractions are only 
about 14\% and 6\% respectively.
However, part of the 
foreground stars have already been identified from the low value of 
their extinction (Sect. 4.1) 
and taken out of the stars further discussed below. Their number  
is 110, i.e. about 8\% of the total.

To summarize, one may expect that the sample of ISOGAL sources further 
discussed below contains about 6\% of foreground disk stars, 6\% of 
background disk stars, and 88\% of central stars within the bulge limits -- 
$\sim$70\% of bulge stars and $\sim$18\% of central disk stars. However, 
these percentages must be considered as indicative since they vary with the 
Galactic coordinates and they result from approximate estimates with a sketchy 
axisymmetrical bulge model.

\section{Luminosities}

After dereddening (Sect. 4.1), the bolometric magnitudes (M$_{\rm bol}$) are 
derived for the stars using the bolometric corrections (BC$_{\rm K}$) and a 
distance 
modulus of 14.5. We have used an approximate relation (Fig. 6) between 
BC$_{\rm K}$ {\it vs} (K$_{\rm s}$-[15])$_{\rm 0}$ to derive the bolometric 
corrections for 
the ISOGAL sources by combining the BC$_{\rm K}$ values derived as follows 
from Frogel \& Whitford (1987) and from Groenewegen (1997). For 
(K$_{\rm s}$-[15])$_{\rm 0}$ $<$ 3, we have used 24 ISOGAL Baade's Windows 
sources 
(shown by open circles in Fig. 6) which are common in 
Frogel \& Whitford's (1987) catalogue with a value for 
BC$_{\rm K}$, to derive the average curve displayed in Fig. 6. We have extended 
this curve for (K$_{\rm s}$-[15])$_{\rm 0}$ $>$ 3 from 
the approximate relation between 
BC$_{\rm K}$ {\it vs} (K-[12])$_{\rm 0}$ for IRAS AGB stars proposed by 
Groenewegen 
(1997). For this purpose, we 
have used an approximate empirical relation between y~=~(K-[12])$_{\rm 0}$ 
and x~=~(K$_{\rm s}$-[15])$_{\rm 0}$ -- namely, y~=~x~-~(x/7)$^{3}$ -- 
 derived from the modelling of infrared emission of nearby 
oxygen-rich AGB stars (Groenewegen, private communication). The 
BC$_{\rm K}$ values from Groenewegen (1997) are shown as open squares in
Fig. 6. However, they have been slightly rescaled by adding 
0.25 mag to match them with  
Frogel \& Whitford's values for (K$_{\rm s}$-[15])$_{\rm 0}$~$\la$~3. 
Finally, a 
polynomial fit has been performed between BC$_{\rm K}$ and 
x~=~(K$_{\rm s}$-[15])$_{\rm 0}$ 
(see caption of Fig. 6). We have checked that such a simple approximate 
determination of M$_{\rm bol}$ agrees within a few tenths of magnitude with an 
elaborate integration of all DENIS and ISOGAL flux densities along the method 
of Loup et al. (in preparation). 

The luminosity of each star is derived from M$_{\rm bol}$ using the classical 
relation, $\rm L (L_{\odot}) = 10^{-(M_{bol} - 4.75)/2.5}$, and is available in 
Table 2. Figs. 7a,b show the histograms of bolometric magnitudes and 
luminosities
of the sources in ISOGAL bulge fields. These distributions are clearly
incomplete below the luminosity of the bulge RGB tip. One should keep
in mind that for $\sim$ 12\% of the sources, the luminosities thus 
derived are wrong because their distances are significantly different
from the bulge distance, either foreground or background on the line
of sight. This will spuriously broaden the luminosity 
distribution. In particular, most of the sources appearing with very
large luminosities are probably foreground stars with smaller 
luminosities. In addition, even for the other sources within the ``bulge'' 
limits (6--10 kpc), the difference in distances with respect to 8 kpc induces 
a spread by about one magnitude (but with an r.m.s.~$\la$~0.2 mag because the 
distribution is very centrally peaked). Other uncertainty terms, to be 
combined in quadrature, include the residuals to the BC$_{\rm K}$ fit and the 
BC$_{\rm K}$ calibration (r.m.s.~$\la$~0.2 mag), uncertainties in the K and 
15 $\mu m$ photometry (r.m.s.~$\la$~0.2 mag), deredddening for large 
extinctions (r.m.s.~$<$~0.2 mag), variability (r.m.s.~$\sim$~0.2 mag for 
strong variables, $<$~0.1 mag for others). The combined r.m.s. uncertainty for 
the luminosity of bulge sources is thus probably less than 0.4--0.5~mag.

\begin{figure}[h]
\centering
\resizebox{\hsize}{!}{\includegraphics{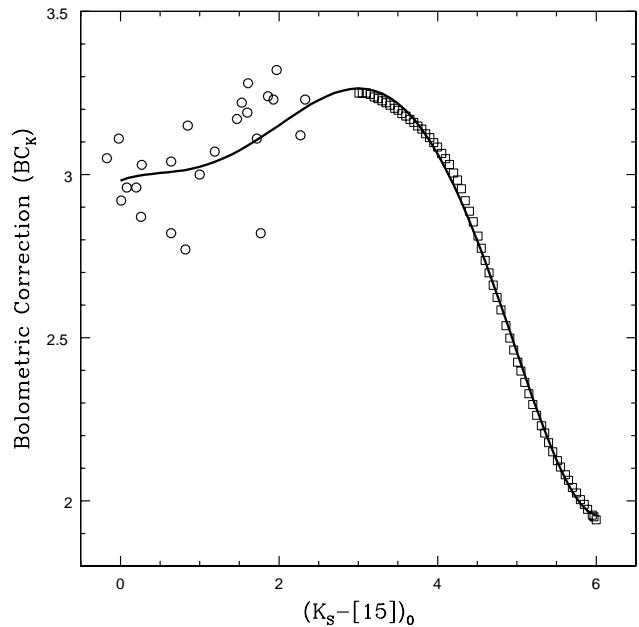}}
\caption[]{Approximate relation between bolometric correction, BC$_{\rm K}$,  
{\it versus} x~=~(K$_{\rm s}$-[15])$_{\rm 0}$ {(BC$_{\rm K}$=2.982+0.0962x--0.1653x$^2$+
0.1499x$^3$--0.04173x$^4$+0.003351x$^5$)}
derived using the BC$_{\rm K}$ values from 
Frogel \& Whitford (1987) (open circles) and Groenewegen (1997) (open squares) 
(see in the text).}
\label{FIG6}
\end{figure}

\begin{figure}[h]
\centering
\resizebox{\hsize}{!}{\includegraphics{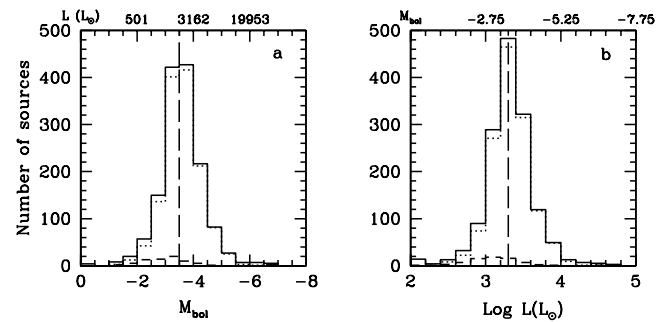}}
\caption{Histograms showing bolometric magnitudes and luminosities of the 
sources in ISOGAL bulge fields. Solid lines show all ISOGAL sources detected
at K$_{\rm s}$ and 15 $\mu m$. Dotted lines show the sources having 
ISOGAL/DENIS association quality flags 4 \& 5. Short-dashed lines show the 
sources having
quality flags 2 \& 3. Long-dashed vertical lines indicate the bulge RGB tip 
(M$_{\rm bol}$ $\sim$ --3.5; L $\simeq$ 2000 L$_{\rm \odot}$).}
\label{FIG7}
\end{figure}

\begin{figure}[h]
\centering
\resizebox{\hsize}{!}{\includegraphics{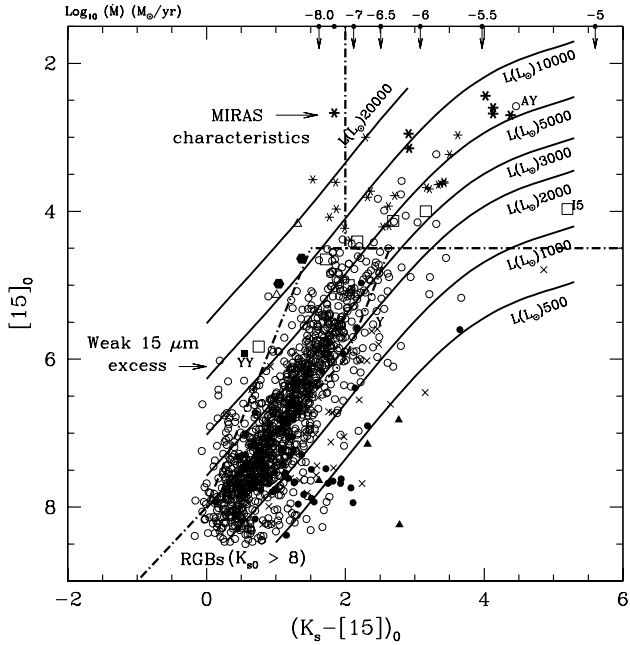}}
\caption{[15]/K$_{\rm s}$-[15] magnitude-colour diagram of ISOGAL sources with 
DENIS/2MASS counterparts. The open circles show the sources having 
ISOGAL--DENIS association quality flags of 4 \& 5. The crosses show the 
sources having association quality flags of 2 \& 3 (see the text). The filled 
hexagons display the ISOGAL sources associated with 2MASS counterparts 
(since the DENIS sources [K$_{\rm s}$ $\le$ 6.5] are saturated, they have been 
replaced by the 2MASS sources). The remaining saturated DENIS sources 
(which have no published 2MASS counterparts) are shown by filled squares. 
The bright ISOGAL sources associated with 2MASS sources (which have no DENIS 
counterparts) are shown by open triangles. Sources without 7 $\mu m$ 
associations are displayed by filled circles. The filled triangles show the 
ISOGAL sources associated with 
2MASS sources but having no DENIS counterparts. The IRAS sources are shown
by open squares. However the four very red IRAS sources I1--4 (Table 3) are 
outside the limits of the figure. The asterisks show the sources 
detected at 8 $\mu m$ in the MSX survey which are
associated with strong ISOGAL sources ([15] $<$ 4.5 mag).
Among them the dark ones are those associated with the MSX D
(15 $\mu m$) sources ([15] $<$ 3.5 mag). The curves from bottom to top 
represent the luminosity estimates from 500 to 20000 L$_{\rm \odot}$, 
respectively 
(see  Fig. 5). Various regions discussed in the text 
(see Sect. 6) are indicated with boundaries as dotted lines. 
The approximate scale of mass-loss rate displayed at the top of the 
figure is derived from Eq. (1) given by Jeong et al. 2002 (see Sect. 7).}
\label{FIG8}
\end{figure}

\section{Nature of the ISOGAL sources}

Figs. 8, 9 \& 11 show the (K$_{\rm s}$-[15])$_{\rm 0}$/[15]$_{\rm 0}$,  
(K$_{\rm s}$-[15])$_{\rm 0}$/K$_{\rm s}$$\rm {_0}$ and [7]-[15]/[15] 
colour-magnitude diagrams of ISOGAL sources, respectively. The first two
diagrams are more directly related to the luminosity and mass-loss 
derivations (Sects. 5 \& 7), and luminosity and 
mass-loss rates estimate are displayed in Figs. 8 \& 9. However, such 
derivations rely on a good extinction correction of 
K$_{\rm s}$ magnitudes. On the other hand, the diagram [7]-[15]/[15], 
which depends 
little on extinction, provides a straightforward view of the nature and the 
amount of dusty sources of various classes. It even allows a direct measure of 
the mass-loss rates, but with less sensitivity, especially for the largest and 
the smallest ones. 

Figs. 8, 9 \& 11 are very similar to the same diagrams in Omont et al. (1999) 
for the field at $l = 0$, $b = +1^\circ$ (see also Glass et al. 1999). One can distinguish three main classes of sources, corresponding to three regions delimited in Figs. 8, 9 \& 11 :

There are first, a number of stars with luminosity 
\mbox{($<$ 2000 L$_{\rm \odot}$)} 
and 
colours \mbox{[(K$_{\rm s}$-[15])$_{\rm 0}$ $<$ 1]} and [7]--[15] consistent 
with dustless 
stars on the RGB close to its tip, which is at K$_{\rm 0}$ $\sim$ 8.0 as 
quoted by 
Frogel et al. (1999)\footnote{As quoted in footnote 2 of Omont et al. (1999), 
we neglect the small difference of \mbox{K--K$_{\rm s}$ $\sim$ 0.04}, for such 
stars.}, which corresponds to \mbox{M$_{\rm bol}$ $\simeq$ --3.5} and 
L $\simeq$ 2000 L$_{\rm \odot}$. 
Their spectral types should be $\sim$M3--M5 (Frogel \& Whitford 1987,  Glass et al. 1999), which corresponds to colour 
\mbox{(K$_{\rm s}$-[15])$_{\rm 0}$ $\simeq$ 0.71--0.78} with 
T$_{\rm eff}$ $\simeq$ 3600--3400 K, 
respectively. 

Then, a whole sequence of sources display luminosities slightly above the 
RGB tip (2000--5000 L$_{\rm \odot}$) and colours with a weak 15 $\mu m$ excess 
[1 $<$ (K$_{\rm s}$-[15])$_{\rm 0}$ $<$2], characteristic of small amounts 
of dust and 
thus of weak mass--loss; they were called ``intermediate AGB'' stars by 
Omont et al. (1999). The K$_{\rm s0}$ magnitude range from the base to 
tip of the 
sequence (Fig. 9) corresponds to M spectral types from about M6 to M8--M9 
(Frogel \& Whitford 1987, Glass et al. 1999). It is known from 
Alard et al. (2001) that they are semi-regular variables with relatively short 
periods and weak amplitudes. 

Finally, less numerous sources, in the top right corner of Figs. 8 and 11 
([15]$_{\rm 0}$ $<$ 4.5, (K$_{\rm s}$-[15])$_{\rm 0}$ $>$ 2, 
\mbox{([7]--[15])$_{\rm 0}$ $\ga$ 1.3)}, may have 
larger luminosities and super--wind mass--loss characteristic of Miras 
(see Glass et al. 1999, Alard et al. 2001, Omont et al. 1999). Most of them 
are detected at 8 $\mu m$ in the MSX survey (Price et al. 2001) as seen in Table 2.

\begin{table*}
\setcounter{table}{2}
\caption[]{Catalog of brightest and reddest ISOGAL sources. 
Data for each source are displayed in two lines, with the ISOGAL standard name
(e.g. ISOGAL-PJ173812.5-293938) and the IRAS name if any. In order to make 
the easy comparison with ISOGAL 7 \& 15 $\mu m$ magnitudes, [7] \& [15], and 
IRAS flux densities (in Jy) at 12 \& 25 $\mu m$, S12 \& S25, MSX intensities 
are given in magnitudes for bands A ($\sim$ 8 $\mu m$) and 
D ($\sim$ 15 $\mu m$), 
and in flux densities (in Jy) for bands C ($\sim$ 12 $\mu m$) and 
\mbox{E ($\sim$ 21 $\mu m$)}.} 
\vspace{0cm}
\hspace*{-1cm}
\begin{tabular}{l c c c c c c c c c c c c c }\\
\hline
Name (ISOGAL--P..)  & $l$  &  J   & K$_{\rm s}$ & [7] & [15] & A$_{\rm V}$ & 
V$_{\rm LSR}$$^1$ & S12 & S25 &  L$^2$ & $\rm \dot {M}$ & Id.\\
IRAS Name     & $b$  &  &   & msxA  & msxD &         &   & msxC  & msxE &  & &   & \\
 & (deg) & (mag) & (mag) & (mag) & (mag) & (mag) & (km/s) & 
(Jy) & (Jy) & 10$^4$ L$_{\rm \odot}$ & (M$_{\rm \odot}$/yr)\\  
\hline    
J173812.5-293938 & -1.5 & 11.75 &  8.16  & 5.26 &  2.89 & 6.3  &       &     &      &  0.32 & 4.9$\times$10$^{-6}$  & AY\\
 & 1.0 & & & & & & & & & & & \\
J174122.7-283146 & -0.1 & $>$14.16 & 11.80 & 3.47 & 1.54 & 8.9 & -51.0 & 2.0 & 4.3  &  0.89   & 3.2$\times$10$^{-5}$ & I3\\
17382-2830  &  1.0 &  &  & 3.69 & 1.87 &     &            & 2.5    & 4.7     &  &     \\
J174458.5-271442 & 1.4  & 12.83    & 10.80 & 0.90 &-0.51  & 6.1 &-12.3  &15.0 & 51.3 & 8.87 &4.1$\times$10$^{-5}$ & $^3$I1\\
17418-2713  & 1.0  &    &  & 1.21 & -1.28 &     &       & 31.2    & 80.3     &      & \\
J174506.1-271516 & 1.4 & 10.92 &  8.67  & 8.43 &  5.84 & 5.4 &  &       &      &  0.16 & 2.9$\times$10$^{-7}$ & Y\\
  & 1.0 & & & & & & & & & & & \\
J174522.9-271109 & 1.5 & $>$15.88 & 12.50 & 2.65 & 1.42 & 8.0 & & 4.5 & 4.9 &  3.67 & 3.6$\times$10$^{-5}$ & I4\\
17422-2709  & 1.0 &    &  & 2.74 & 1.23 & &   & 4.6    & 5.7    &  & \\
(J174533.2-304943) & 1.6 & (11.13)   & (7.64)  & (6.66) & (6.13) & (11.9) &  &  & 1.6: & 0.66 & & YY \\  
17423-3048  & -1.0 &         &       & (6.07) &      &      &      & \\ 
J174548.6-304312 & -1.5 & $>$15.26 & 10.14  & 4.38 &  1.58 & 12.3 & & 3.4 & 5.3  &  0.17 & 1.7$\times$10$^{-5}$ & I2\\
17426-3042  & -1.0 &   &   & 3.67 &  1.85 & & & 2.7 & 4.2 & & \\
J174939.4-292723 & 0.0 & 12.31 &  9.85  & 6.34 &  4.16 & 7.6 & & & 2.5 &  0.12& 8.1$\times$10$^{-6}$ & I5\\
17464-2926 & -1.0 & & & & & & & & & & & \\
\hline
\end{tabular}
\vspace *{0.2cm}
\hspace {-0.5cm}
$^1$~value comes from OH maser detection; 

\vskip -0.3cm
\hspace {-0.6cm}
~$^2$~if one assumes a distance of  8~kpc; 

\hspace {-0.65cm}
~$^3~$the ISOGAL magnitudes and M$_{\rm bol}$ of I1 are uncertain because the ISOGAL flux densities are 

\hspace {-0.32cm}
above saturation limits of the detectors.
\end{table*}

\subsection {Brightest and Reddest ISOGAL sources}

We have identified 16 IRAS sources within the fields of Table 1. They are 
associated with 15 bright ISOGAL stars ([15]$_{\rm 0}$ $\la$ 6.0). Table 3 
shows the brightest and reddest ISOGAL sources with IRAS and MSX associations 
when available.
These sources are shown by open squares in Figs. 8, 9, 11 \& 12. Out of these 
15 sources, the four brightest ones ([15]$_{\rm 0}$ $<$ 1.5) are denoted 
by ``I1--4'' 
in Table 3 and Figs. 9 \& 12. These four sources visible in 
Figs. 9 and 11 are not shown in Figs. 8 \& 12. All of them have extremely red values of
(K$_{\rm s}$--[15])$_0$ characteristic of large amount of dust. 
Their [12]--[25] IRAS 
colours are characteristic of AGB stars and not of young stellar objects 
or post-AGB stars. They are thus probably AGB stars with very 
large mass-loss rates, $>$~10$^{-5}$ M$_{\rm \odot}$/yr 
(see Sect. 7 and Table 3). 
Table 3 displays approximate values of their bolometric fluxes computed 
from an integration of their available flux densities, either following the 
prescriptions of Loup et al. (in preparation) or from a more approximate 
integration when (K$_{\rm s}$--[15))$_{\rm 0}$ is too large to use Loup et al's method. 
All of them correspond to large values of bolometric luminosities if the 
sources are at the Galactic Center distance. Such luminosity values are 
consistent with the very large mass-loss rates and both quantities point out 
to relatively large values of the initial mass. However, it is not clear 
whether these stars belong to the central bulge/disk or to the foreground disk 
(the four of them have $|$$b$$|$ $\simeq$ 1). Their large intrinsic colours 
make it very difficult to determine their interstellar extinction. 
Their following individual properties may help to decide about their 
bulge membership: 

	-- {\bf I1} -- It is the brightest 15 $\mu m$ source and its ISOGAL 
magnitudes are uncertain because of saturation. It belongs to the most 
extreme class of OH/IR stars with a 
massive cold circumstellar envelope as shown by its exceptional [12]--[25] 
colour and strong 10 and 18 $\mu m$ absorption bands in the IRAS LRS spectrum
(IRAS Science Team 1986). The radial velocity and the very large bolometric 
flux point to a foreground disk object.

	-- {\bf I2} -- The bolometric flux favours bulge membership.

	-- {\bf I3} -- It is also an OH maser, but its radial velocity and 
relatively small bolometric flux point to bulge membership.

	-- {\bf I4} -- The bolometric flux and the blue [7]--[15] colour 
favour foreground disk membership.

	Three other stars with very red [7]--[15] colours (Fig. 11) are also 
listed in Table 3. Their nature -- AGB, post--AGB, PN, YSO -- is uncertain, 
and they are not detected by MSX. We have also added an IRAS source, YY, 
with no ISOGAL point source counterpart:

	 -- {\bf I5} -- From its position in the [7]/[7]--[15] diagram, it 
could be either a young stellar object (YSO) or an AGB star 
(Schuller et al. in preparation). However, its 
non detection at 12 $\mu m$ together with the large value of the ratio 
S$_{\rm 25}$$_{\mu m}$/S$_{\rm 15}$$_{\mu m}$ $\sim$ 6, and its small 
bolometric flux 
could indicate a young stellar object (or a planetary nebula).

	-- {\bf AY} -- Again, from its position in the [7]/[7]--[15] and 
K$_{\rm s0}$/(K$_{\rm s}$--[15])$_{\rm 0}$ diagrams, and from its bolometric 
flux, it could be a YSO or an AGB star. 

	-- {\bf Y} -- It could appear as a good 
YSO candidate. However, its ISOGAL magnitudes could have been affected by the 
track of the saturated star I1 in the ISOGAL images. Its completely anomalous 
colour (K$_{\rm s}$-[7])$_{\rm 0}$ = --0.15 also points to a wrong 
7 $\mu m$ magnitude.

	-- {\bf YY} -- This source, IRAS17423--3048, is probably a massive 
YSO from its 
IRAS 25--60 $\mu m$ flux densities. It has no convincing association in the 
\mbox{ISOGAL-DENIS} PSC, although the ISOGAL--DENIS--MSX source displayed 
within bracketts in Table 3 is within the limits of a possible association. 
On the other hand, there is some extended emission at its position 
in the 15 $\mu m$ ISOGAL image.

\subsection{Other detected sources}

All other IRAS detected sources have \mbox{[12]--[25]} colours indicative of 
AGB stars. The fact that there are practically no other stars in Fig. 11 
meeting the 
criteria \mbox{([7]--[15] $>$ 1.8} with [15]$<$4.5) for credible young star 
candidates defined by 
Felli et al. (2000, 2002), is not surprising in such high galactic 
latitude lines of sight. 

As quoted in Glass (1999), much of the scatter \mbox{($\sim$ 1 mag)} seen in the 
``AGB sequence'' of Figs. 8 \& 11) may be due to the distribution in the 
depth of 
the bulge and the central disk (Glass et al. 1995). Furthermore, in Figs. 8 and 9, it seems unlikely
that there is in addition an anomalously large number of stars far above this
sequence by up to $\sim$ 1.5 mag. Such stars with anomalously bright  
ISOGAL magnitudes with respect to their colours, appear more clearly in 
the [7]-[15]/[15] 
colour/magnitude diagram of Fig. 11. It seems that most of them could not be 
bulge 
AGB stars with very high luminosity, because of the age of the bulge population
and of the low mass-loss rates deduced for most of them from 
(K$_{\rm s}$-[15])$_{\rm 0}$ 
or [7]-[15]. Most of them are probably foreground AGB stars with moderate
luminosity and mass-loss. Their population and the distribution of the values 
of the 15 $\mu m$ magnitude excess with respect to the bulge sequence look in 
rough agreement with the proportion and the distance distribution expected for 
the disk foreground stars (Sect. 4). Similarly, most sources which lie below 
the main bulge AGB sequence in Figs. 8, 9 \& 11, are likely background disk
AGB stars.

There are a few stars (Fig. 9) with large values of 
\mbox{K$_{\rm s}$-[15])$_{\rm 0}$ ($>$ 1)} 
below the RGB tip. But their case is not clear. Only a few of them have values 
of ([7]--[15])$_{\rm 0}$ $>$ 0.6. Most of the sources have poor photometry 
because 
they are close to the 15 $\mu m$ detection limit or of poor association quality. 
In the following discussion, we will drop the most doubtful cases, i.e. all 
stars with K$_{\rm s}$${_{\rm 0}}$ $>$ 9.0 (Fig. 9). The nature of the few 
remaining cases 
is unclear. A number could be background stars with large mass-loss. 
Additional observations are required to confirm the nature of these sources 
and to exclude the possibility of spurious associations with the K$_{\rm s}$ 
sources. However, such a 15 $\mu m$ excess could be an indication that some 
RGB stars close to the RGB tip have a significant mass-loss, 
as found in globular clusters by Origlia et al. (2002).

The 28 sources detected only at 15 $\mu m$ (Sect. 3) are absent from all 
colour--magnitude diagrams. However, they remain candidates for YSOs or 
Planetary Nebulae, especially for half of them with relatively good quality 
flags. The strongest one, \mbox{ISOGAL--PJ1753178--280433}, \mbox{[15] = 5.88},
is identified in SIMBAD as a Planetary Nebula JaSt 90 
(Van de Steene \& Jacoby 2001).

Fig. 12 shows [7]/K$_{\rm s}$-[7] magnitude-colour diagram of ISOGAL sources. 
The [7]/K$_{\rm s}$-[7] diagram seems to be the best criterion for the 
detection of 
large amplitude LPVs (Glass et al. 1999, Schultheis et al. 2001). The region 
of high mass-loss AGB stars 
\mbox{((K$_{\rm s}$-[7])$_{\rm 0}$ $>$ 1}, [7]$_{\rm 0}$ $<$ 7) is 
explicitly drawn in this diagram.  

\begin{figure}
\centering
\resizebox{\hsize}{!}{\includegraphics{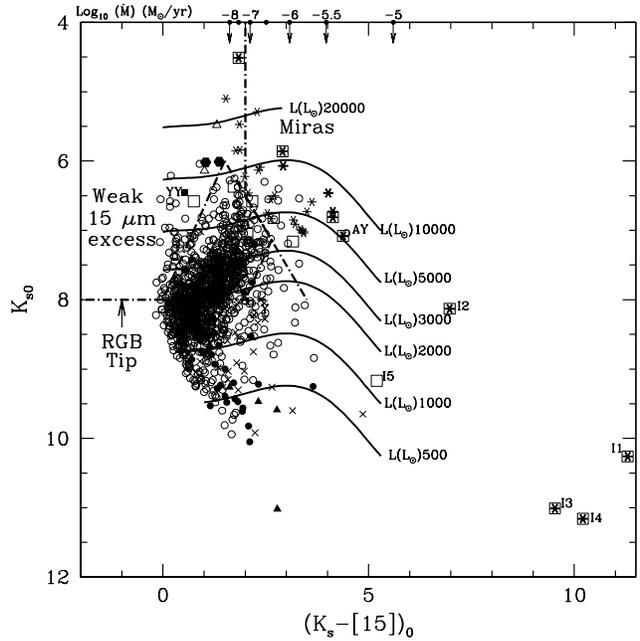}}
\caption{K$_{\rm s0}$/(K$_{\rm s}$-[15])$_{\rm 0}$ magnitude-colour diagram 
of ISOGAL sources
with DENIS/2MASS counterparts. The symbols and the curves are same as shown in 
Fig. 8. The red IRAS sources of Table 3 are denoted by ``I'' (see text). They 
are also MSX D sources associated with the strong 15 $\mu m$ sources. 
The RGB tip (K$_{\rm s0}$ $\sim$ 8.0) is shown as dotted line. Other two
regions are same as in Fig. 8. 
}
\label{FIG9}
\end{figure}

\begin{figure}
\centering
\resizebox{\hsize}{!}{\includegraphics{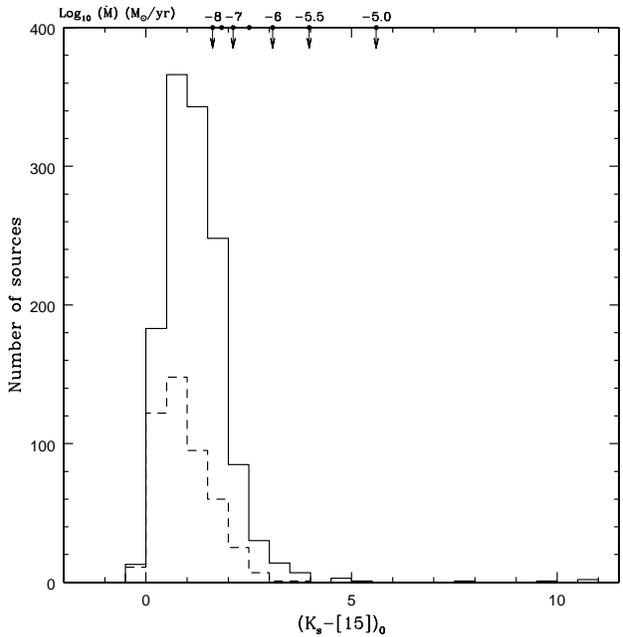}}
\caption{K$_{\rm s}$-[15] colour distribution of ISOGAL sources. The mass-loss 
scale at the 
top of the figure are same as shown in Fig. 8. The full line histogram 
represents the stars of all nine fields of our sample (Table 1). The dashed 
line represents stars in the fields with $|$$b$$|$ $\ge$ 2$^\circ$. 
}
\label{FIG10}
\end{figure}

\begin{figure}
\centering
\resizebox{\hsize}{!}{\includegraphics{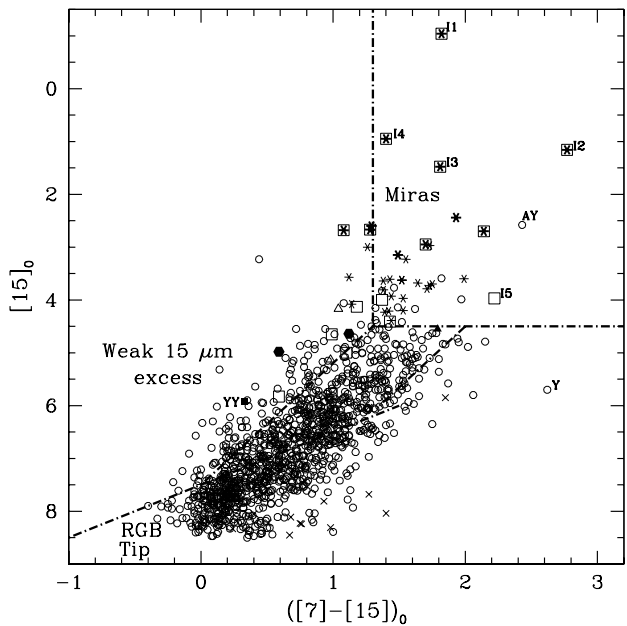}}
\caption{[15]/[7]-[15] magnitude-colour diagram of ISOGAL sources detected in 
the ISOGAL fields. The open circles show the ISOGAL sources 
having 7/15 $\mu m$ association quality flags of 3 \& 4. The crosses show the 
sources having association quality flags of 2. The filled hexagons display the 
associations
with 2MASS (since the DENIS sources [K$_{\rm s}$ $\le$ 6.5] are saturated, 
they
have been replaced by 2MASS). The remaining saturated DENIS sources 
(which have 
no 2MASS counterparts) are shown by filled squares. The bright ISOGAL sources
associated with 2MASS sources (which have no DENIS counterparts) are shown
by open triangles. 
The filled triangles show the 
ISOGAL sources which have no available DENIS nor 2MASS counterparts,
mostly because of saturation or non observation. 
The asterisks show the sources detected at 8 $\mu m$ in the 
MSX survey which are associated with strong ISOGAL sources ([15] $<$ 4.5 mag). 
Among them the dark ones are those associated with the MSX D (15 $\mu m$)
sources ([15] $<$ 3.5 mag). The IRAS sources are shown
by open squares and the brightest ones are denoted as ``I'' (see text).
Sources with larger luminosities and super-wind mass-loss
characteristic of Miras are inside the boundaries shown as dotted lines.
Other two regions are same as in Fig. 8.
}
\label{FIG11}
\end{figure}

\begin{figure}
\centering
\resizebox{\hsize}{!}{\includegraphics{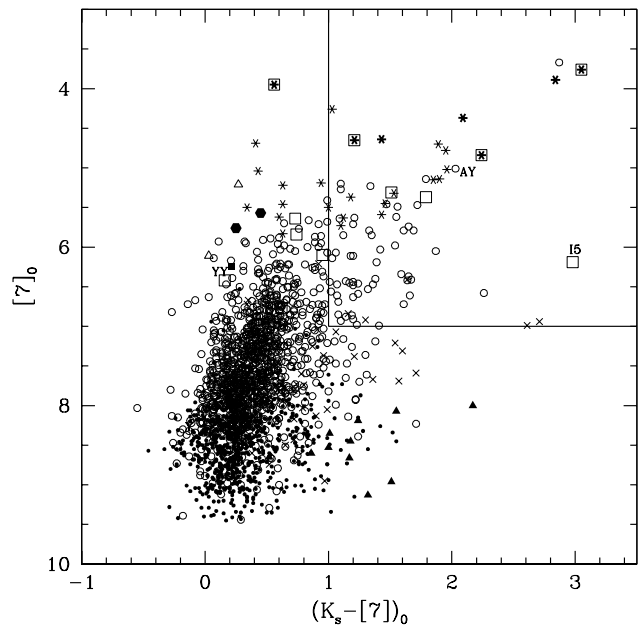}}
\caption{[7]$_{\rm 0}$/(K$_{\rm s}$-[7])$_{\rm 0}$ magnitude-colour diagram 
of ISOGAL sources
with DENIS/2MASS counterparts. The open circles show the sources having 
ISOGAL-DENIS association quality flags of 4 \& 5. The crosses show the sources 
having association quality flags of 2 \& 3 (see the text). The filled hexagons 
display the ISOGAL sources associated with 2MASS counterparts (since the 
DENIS sources (K$_{\rm s}$ $\le$ 6.5) are saturated, they have been replaced
by the 2MASS sources). The remaining saturated DENIS sources (which have
no 2MASS counterparts) are shown by filled squares. The bright ISOGAL sources 
associated with 2MASS sources (which have no DENIS counterparts) are shown
by open triangles. Sources without 15 $\mu m$ associations are displayed by
filled circles. The filled triangles show the ISOGAL sources associated with
2MASS sources but having no DENIS counterparts. The asterisks show the sources
detected at 8 $\mu m$ in the MSX survey which are associated with 
strong ISOGAL sources ([15] $<$ 4.5 mag).
Among them the dark ones are those associated with the MSX D (15 $\mu m$) 
sources ([15] $<$ 3.5 mag). Stars I1--4 are 
outside of the figure limits with (K$_{\rm s}$--[7])$_{\rm 0}$ between 4 and 10 
(see Table 3).
The region of high mass-loss AGB stars 
((K$_{\rm s}$-[7])$_{\rm 0}$ $>$ 1, [7]$_{\rm 0}$ $<$ 7) is drawn in the 
upper right corner (see Fig. 9 of Omont et al. 1999).  
}
\label{FIG12}
\end{figure}

\section{Mass-loss rate ($\rm \dot{M}$) determination}

It is generally agreed that the colour (K-[12])$_{\rm 0}$ 
(where [12] is the magnitude 
corresponding to the IRAS 12 $\mu m$ flux density) is a good indicator of 
the mass-loss 
rate $\rm \dot{M}$ of AGB stars (see e.g. Whitelock et al. 1994, 
Le Bertre \& Winters 1998, Jeong et al. 2002). It is clear that 
(K$_{\rm s}$-[15])$_{\rm 0}$ 
should be an equally
good gauge of $\rm \dot{M}$, with a slightly different calibration. 
However, the relation between $\rm \dot{M}$
and (K$_{\rm s}$-[15])$_{\rm 0}$ or (K-[12])$_{\rm 0}$ directly depends 
on the nature of dust and
on the gas-to-dust ratio. For instance, there is evidence from the preliminary analysis of
the ISOGAL CVF data of a few of these AGB bulge sources that their dust is
dominated by alumina for low and intermediate mass-loss rates  (Blommaert et al. 2001 and in preparation). 
In addition to further analysis and observations of mid--IR spectra, an 
essential element for the accurate calibration of the relation between 
$\rm \dot{M}$
and (K$_{\rm s}$-[15])$_{\rm 0}$ is the
detailed theoretical analysis of mass-loss physics, dust formation and 
radiative transfer, such as performed 
by Jeong et al. (2002 and in preparation). The latter is based on a consistent
time dependent treatment of hydrodynamics, thermodynamics, equilibrium
chemistry and dust formation. Such a theoretical modelling is presently
the only way to infer the gas-to-dust ratio and the mass-loss rates for such
bulge sources in the absence of any direct measurement of the mass of
circumstellar gas and even of the expansion velocity. As they quote, 
the relation derived between $\rm \dot{M}$ 
and (K-[12])$_{\rm 0}$ by Jeong et al. (2002) gives substantially higher 
mass-loss rates than the empirical relation of Le Bertre and Winters (1998) 
(Fig. 13). On the other hand, the agreement is better with the empirical 
values of Whitelock et al. (1994) derived from the classical prescription of 
Jura (1987) giving $\rm \dot{M}$ from far-infrared emission (Fig. 13). 
Here, we will apply the explicit relation between $\rm \dot{M}$ 
and (K-[15])$_{\rm 0}$ derived by Jeong et al. (2002), 

\begin{equation}
	{\rm log~\dot{M} = -6.83/(K-[15])_0 - 3.78}
\end{equation}

with $\rm \dot{M}$ in M$_{\rm \odot}$/yr (Fig. 13). 

Alternatively, one could also infer $\rm \dot{M}$ from 
(K$_{\rm s}$-[7])$_{\rm 0}$ or from [7]-[15].
Deriving $\rm \dot{M}$ from [7]-[15] (Alard et al. 2001) has the advantage 
that the latter quantity
depends very little on the extinction. However, this method is less sensitive
since [7]-[15] varies little. 
On the other hand, (K$_{\rm s}$-[7])$_{\rm 0}$ has a low sensitivity for
small mass-loss rates, but could be conveniently used for large ones.

It is important to notice that there may also be an appreciable uncertainty 
in the
determination of mass-loss rates from (K-[15])$_{\rm 0}$ for strongly 
variable stars where we have only 
single epoch measurements of K$_{\rm s}$ and [15], and at different epochs for 
K$_{\rm s}$ and [15] (Table 1). As quoted, this uncertainty is reduced by 
taking 
the average of DENIS and 2MASS K$_{\rm s}$ values and of ISOGAL and 
MSX 15~$\mu m$ 
magnitudes. One can thus estimate that the global r.m.s. uncertainty on 
(K$_{\rm s}$-[15])$_{\rm 0}$ remains smaller than 0.3--0.4 mag, except for 
the weakest 
sources (K$_{\rm s}$${_{\rm 0}}$ $>$ 9.0, see Sect. 6.2). However, because of 
the rapid 
variation of $\rm \dot{M}$ {\it versus} (K$_{\rm s}$-[15])$_{\rm 0}$ (Fig. 13), the 
corresponding observational uncertainty on  $\rm \dot{M}$ ranges between a 
factor $\sim$2 for its largest values $>$ 10$^{-6}$ M$_{\rm \odot}$/yr, and 
$\sim$5 for the smallest ones $<$ 10$^{-7}$ M$_{\rm \odot}$/yr. Such values 
do not 
include the model uncertainty in the relation between $\rm \dot{M}$ and 
(K$_{\rm s}$-[15])$_{\rm 0}$.
 
\begin{figure}
\centering
\resizebox{\hsize}{!}{\includegraphics{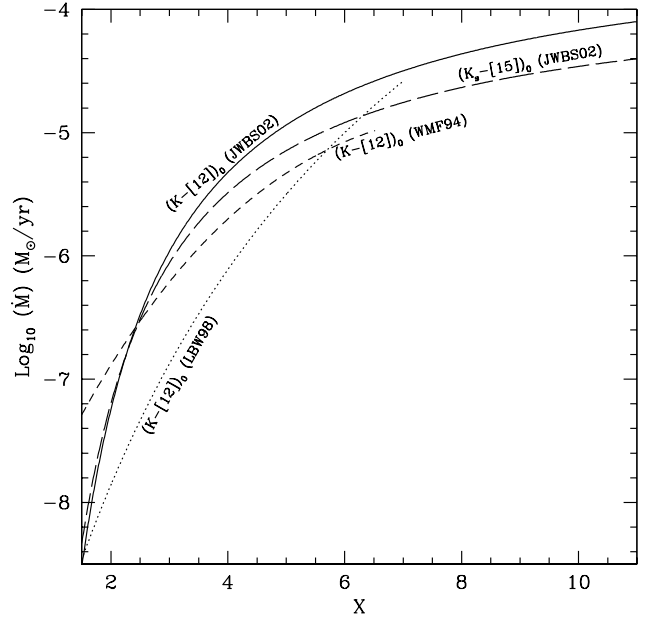}}
\caption{Mass-loss rate ($\rm \dot{M}$) as a function of colour. For  
\mbox{X = (K$_{\rm s}$-[15])$_{\rm 0}$}, Eq. (1) recommended by Jeong et al. 
(2002, JWBS02) is 
used. For X = (K-[12])$_{\rm 0}$, the three curves drawn are derived from 
Jeong et al. (2002) 
with \mbox{log$_{\rm 10}$ $\rm \dot{M}$  = -7.69/(K-[12])$_{\rm 0}$) - 3.40}, 
Whitelock et al. (1994, WMF94) and Le Bertre \& Winters (1998, LBW98), 
respectively.
} 
\label{FIG13}
\end{figure}

\section{Total mass-loss rate in the bulge}

Fig. 14a shows the number distribution of mass-loss rates ($\rm \dot{M}$) of 
ISOGAL sources in the bulge as a function of $\rm \dot{M}$ from Eq. (1) 
along models of Jeong et al. (2002). Fig. 14b displays the corresponding 
values of the average total mass-loss rate per square degree and per 0.5 bin 
of log$~\rm \dot{M}$. It should be remembered that the number of sources is 
incomplete for 
\mbox{$\rm \dot{M}$ $\la$ 10$^{-7}$ M$_{\rm \odot}$/yr}. It is not excluded 
that still lower mass-loss rates, 
$\rm \dot{M}$ $<$ 10$^{-8}$ M$_{\rm \odot}$/yr, contribute a little to 
the integrated mass-loss in the bulge, because of the very large number of 
stars with 1 $\la$ (K$_{\rm s}$-[15])$_{\rm 0}$ $\la$ 2 (Fig. 10), where 
such 15 $\mu m$ 
excess, if real, still shows the presence of circumstellar dust. 

However, because of the very steep distribution of the number of sources 
{\it versus} (K$_{\rm s}$-[15])$_{\rm 0}$ (Fig. 10) and the rapid variation of 
$\rm \dot{M}$ {\it versus} (K$_{\rm s}$-[15])$_{\rm 0}$ (Fig. 13), the error 
in \mbox{(K$_{\rm s}$-[15])$_{\rm 0}$} due to strong variability 
makes the contribution of the corresponding bins larger than in reality. 
Therefore, in Figs. 14b \& d, we have reduced the contributions of the three 
bins between 10$^{-7}$ and 3~10$^{-6}$ M$_{\rm \odot}$yr$^{-1}$ by factors 
1.4, 1.2 and 1.1, respectively, as shown in Figs. 14a \& c. However, we have 
not corrected the bins below 
\mbox{10$^{-7}$ M$_{\rm \odot}$yr$^{-1}$}, 
because the incompleteness of detections of 
such stars could compensate for this effect. 

It is difficult to fully appreciate the uncertainty in the data of 
Figs. 14a \& 14b. There are various sources of observational uncertainty: 
errors in K$_{\rm s}$-[15])$_{\rm 0}$ and their uncertain correction, 
bulge membership, 
Poisson error due to the small number of sources with large $\rm \dot {M}$, 
incompleteness of detections of stars with small $\rm \dot {M}$, etc. They 
could 
combine to give a substantial error, but probably smaller than a factor two. 
However, the major source of error is probably still the calibration of the 
relation between \mbox{K$_{\rm s}$-[15])$_{\rm 0}$ \& $\rm \dot {M}$}, which depends 
on dust 
properties and the dust--to--gas ratio. Its modelling depends on complex 
stellar properties and remains difficult despite the spectacular progress 
achieved by Jeong et al. (2002). 

It is seen in Fig. 14b that the contribution to the
integrated mass-loss per bin of log $\rm \dot {M}$ is 
dominated by mass-loss rates larger than 10$^{-6}$ M$_{\rm \odot}$/yr.  
It is even possible  
that the four stars I1--4 
with (K$_{\rm s}$--[15])$_{\rm 0}$ $>$ 6, 
$\rm \dot {M}$  $\ga$ 10$^{-5}$ M$_{\rm \odot}$/yr 
(Table 3), contribute for about half the total. But their 
belonging to the bulge is not fully established (see Sect. 6).

\begin{table}
\setcounter{table} {3}
\centering
\caption[]{Integrated mass-loss (M$_{\rm \odot}$/yr/deg$^2$) in the 
``intermediate''  bulge, from 
Eq. (1) (Jeong et al. 2002) with $\rm \dot{M}$ $>$ 10$^{-8}$ M$_{\rm \odot}$/yr}
\begin{tabular}{l c c }\\
\hline
  &  with I1-4$^\dagger$  &  without I1-4$^\dagger$  \\
\hline
{\it All fields} & 8.1$\times$10$^{-4}$ & 3.7$\times$10$^{-4}$\\
\hline
{\it Fields with $|$$b$$|$ $>$ 2$^\circ$} & 1.2$\times$10$^{-4}$ & 1.2$\times$10$^{-4}$ \\
\hline
{\it Fields with $|$$b$$|$ $<$ 1.5$^\circ$} & 1.4$\times$10$^{-3}$ & 5.9$\times$10$^{-4}$  \\
\hline
\end{tabular}

\noindent
$^\dagger$Stars with $\rm \dot {M}$  $\ga$ 10$^{-5}$ M$_{\rm \odot}$/yr 
 whose bulge membership is 

\hskip -3.6cm uncertain (Table 3 \& Sect. 6.1)
\end{table}

Numerical values for the integrated mass-loss are displayed in Table 4. 
First are values for mass-loss per square degree averaged over all nine 
fields of Table 1, with and 
without including the contribution of the four stars I1--4, respectively. Since 
it is difficult to decide whether such stars belong to the central bulge/disk 
or not (Sect. 6.1), we also give the same quantities for the three fields with 
\mbox{$|$$b$$|$ $>$ 2$^\circ$} where such stars with very large values of 
\mbox{$\rm \dot {M}$ ((K$_{\rm s}$--[15])$_{\rm 0}$ $>$ 4)} 
are absent (see Fig. 10). The 
corresponding 
distribution of mass-loss rates are displayed in Figs. 14c and 14d. In the 
absence of stars with \mbox{(K$_{\rm s}$--[15])$_{\rm 0}$ $>$ 4}, 
it is seen that 
the relative contribution of large mass-loss rates to the integrated 
mass-loss seems smaller. However, the total number of relevant stars with 
(K$_{\rm s}$--[15])$_{\rm 0}$ $>$ 2 is about four times smaller in the 
three fields with $|$$b$$|$ $>$ 2$^\circ$ than in the totality of 
the fields. Therefore, while there are only four stars with 
$\rm \dot {M}$ $>$ 10$^{-5}$ M$_{\rm \odot}$/yr in the totality of the 
fields, one cannot consider that their  
absence in fields with $|$$b$$|$ $>$ 2$^\circ$ is statistically significant.

Such results seems to show that the relative contribution of stars with low 
mass-loss rates 
($\rm \dot {M}$ $<$ 10$^{-6}$ M$_{\rm \odot}$/yr) is more important 
in the mass return to 
the interstellar medium for the old bulge population, which is more 
predominant at $|$$b$$|$ $>$ 2$^\circ$. Even the stars
with $\rm \dot {M}$ $<$ 10$^{-7}$ M$_{\rm \odot}$/yr may have a 
significant contribution. 

It is not clear whether the stars with very large mass-loss rates, 
$\rm \dot {M}$  $\ga$ 10$^{-5}$ M$_{\rm \odot}$/yr, which are probably 
younger with a 
relatively large initial mass, may have an essential contribution, as it seems 
the case for the solar neighbourhood (see e.g. Le Bertre et al. 2001). 
It would be interesting to make a detailed comparison with the solar 
neighbourhood. However, in addition to the question of bulge membership of 
stars with such large mass-loss in our sample, it is difficult to make a 
precise 
comparison with the recent work of Le Bertre et al. (2001) because their 
sample is not as complete for stars with low mass-loss as for large mass-loss 
ones.

We may assume the value of Eq. (A.8) for an approximate average
bulge projected mass density at the average position $l$ = 0$^\circ$, 
$|$$b$$|$ = 2$^\circ$ for our fields, from the model of 
Wainscoat et al. (1992) --
$\Sigma _{\rm Bu}$(0,2) $\simeq$ 5.7$\times$10$^7$ M$_{\rm \odot}$ deg$^{-2}$. 
We infer a rough value for the total (bulge + disk) projected mass density of 
$\simeq$ 8$\times$10$^7$ M$_{\rm \odot}$ deg$^{-2}$, by assuming the same 
disk/bulge ratio, 30/70, as for the stellar counts estimated in the 
conclusion of Sect. 4.2.  
The derived integrated mass-loss rate per unit stellar mass is thus about  
0.4--1.0 10$^{-11}$ yr$^{-1}$.

One notes that there is a difference by a large factor in the average 
mass-loss per square degree between the fields with $|$$b$$|$ $>$ 2$^\circ$ 
and those with $|$$b$$|$ $<$ 1.5$^\circ$. This factor is 5 and 12, without 
and with the contribution of stars I1--4, respectively (Table 4). On the 
other hand, since the bulge projected mass density varies as 
[$b^2$+($\ell$/1.6)$^2$]$^{-0.4}$ from the model of 
Wainscoat et al. (1992) 
(see Eq. A.7)), the corresponding ratio for the average projected mass density 
is only $\simeq$~2 for the bulge. It should be  $\simeq$~3 when one takes into 
account the disk contribution. A possible explanation is that the fields 
closer to the Galactic Center contain a larger proportion of younger stars 
than in the outer bulge (Frogel et al. 1999), which, in particular, contribute 
more to the large mass-loss rates. However, a deeper discussion, with more 
elaborate Galactic models and a better identification of foreground sources, 
will be necessary to well establish and quantify these properties.

\begin{figure*}
\includegraphics[width=17cm]{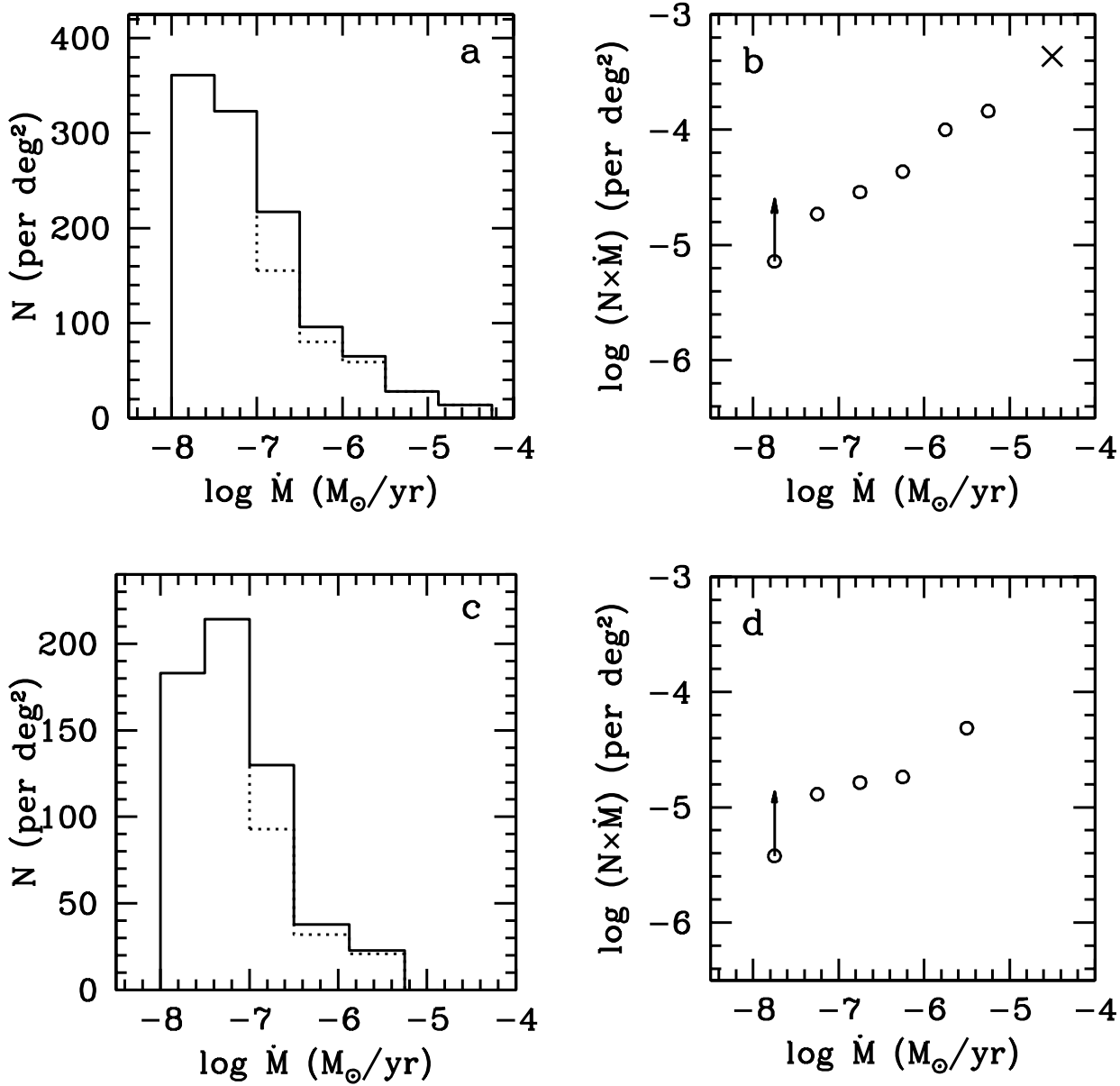}
\caption[]{
(a). Distribution of mass-loss rates ($\rm \dot {M}$) of ISOGAL sources
inferred from (K$_{\rm s}$-[15])$_{\rm 0}$ in the nine bulge fields, 
derived from the 
formula of Jeong et al. (2002), Eq. (1). The dotted line includes a tentative 
correction for the spurious counts due to the error in 
(K$_{\rm s}$-[15])$_{\rm 0}$ and the 
very steep count distribution, as described in the text. 
(b). Average total mass-loss rate per square degree and per 0.5 bin of 
log ($\rm \dot {M}$) as a function of $\rm \dot {M}$ after this correction. 
The contribution of stars I1--4, 
$\rm \dot {M}$  $\ga$ 10$^{-5}$ M$_{\rm \odot}$/yr, is shown by a cross. 
The arrow 
shows the incompleteness on the data for the smallest mass-loss rates. 
(c) \& (d). Same as a \& b, but limited to the three fields with $|b|$ $>$ 2. 
}
\label{FIG14}
\end{figure*}

\section{Conclusion}

We have analysed in detail the AGB population in the available ISOGAL 
``intermediate'' bulge fields ($|$$l$$|$ $<$ 2$^\circ$, 
$|$$b$$|$ $\sim$ 1$^\circ$ -- 4$^\circ$). We have confirmed that the 
combination of near-infrared and 
mid-infrared (7 and 15 $\mu m$) ISOGAL data allows reliable detection of AGB 
stars above the RGB tip with a determination of their luminosity 
(providing that they belong to the bulge) and  
mass-loss rate. We conclude that almost all the $\sim$1300 ISOGAL sources 
detected both at 
7 and 15 $\mu m$ on the line of sight of the bulge are 
AGB stars or RGB tip 
stars. A large proportion of these AGB stars have appreciable mass-loss rates, 
as shown by the excess in (K$_{\rm s}$-[15])$_{\rm 0}$ and 
([7]-[15])$_{\rm 0}$ colours, characteristic of circumstellar dust emission. 
We have performed a preliminary determination of mass-loss rates from 
\mbox{(K$_{\rm s}$-[15])$_{\rm 0}$} from recent theoretical modelling by 
Jeong et al. (2002) which are in agreement with 
Whitelock et al. (1994) for (K-[12])$_{\rm 0}$. 
For the bulk of our sample, and especially at $|$$b$$|$ $\sim$ 1$^\circ$, 
the total mass returned to the interstellar medium is dominated by mass-loss 
rates larger than 10$^{-6}$ M$_{\rm \odot}$/yr. However, four stars, out of 
$\sim$2000, have mass-loss rates $\ga$ 10$^{-5}$ M$_{\rm \odot}$/yr. They 
could dominate the mass return to the interstellar medium if they all 
belonged to the bulge, which is still unclear. In the more outer bulge, 
$|$$b$$|$ $>$ 2$^\circ$, the contribution of large mass-loss rates seems 
less important.

\appendix

\section{Numbers of foreground and background disk sources}

In order to derive an order of magnitude of the proportion of 
foreground and background disk stars, on the line of sight of the observed 
bulge fields, one can use any reasonable simple Galactic model for the 
distribution of contributing stars, e.g. the one by Wainscoat et al. (1992). 
We have quoted that practically all bulge sources detected by ISOGAL are 
M giants later than about M3. Background detected stars should be even 
later M giants, while the great majority of foreground disk sources are K 
or early M giants. It would be straightforward, but tedious, to determine 
the proportion of 
foreground and background disk stars in any given magnitude bin, on various 
lines of sight, by  direct numerical integrations of the Galactic 
distributions of the various classes of stars -- especially red giants -- of 
Wainscoat et al. However, we take advantage of the simple behaviour of the 
giant luminosity function to make an approximate analytical integration. 
The K--M giant luminosity function is 
roughly proportional to L$^{-1}$ (see e.g. Frogel \& Whitford 1987).
At the point (r,z) in the direction ($\ell$, $b$), with $b$~$<<$~1 and 
z~=~r$b$, the number of such giants per unit volume and absolute magnitude 
bin dM$_i$ is $\rho _{g}$(r,z)~$\times$~f$_i$(M$_i$)dM$_i$ where 
f$_i$(M$_i$) is approximately proportional to 10$^{\rm M_i/2.5}$. The 
number of 
stars per unit volume at point (r,z) contributing to the observed magnitude 
bin dm$_i$, with M$_i$~=~M$_i$~--~2.5log(r/10pc)$^2$, is thus 
proportional 
to r$^{-2}$~$\times$~$\rho _{g}$(r,z). The number of counts per unit of solid 
angle $\Omega$ (e.g. deg$^2$) in a magnitude bin in the direction 
($\ell$, $b$), is then proportional to 

\begin{eqnarray*}
	\int{r^{-2} \, \rho _{g}(r,z) \, \Omega \, r^2 dr}  \propto   \int{\rho _{g}(r,z)~ dr} \hskip 2.4cm (\rm A.1)
\end{eqnarray*}

We have therefore integrated the density models of Eqs. (4) and (5) of 
Wainscoat et al. (1992) for the disk and the bulge, respectively, on a 
typical line
of sight for our fields (e.g. $l = 0$, $b = 2^\circ$). For the disk, 

\begin{eqnarray*}
	\rho _{gD} (r,z)  = \rho _{0g}\, exp{[-(R-R_0)/h\, - \, |z|/h_z]} \hskip 1.4cm (\rm A.2)
\end{eqnarray*}

where $\rho _{0g}$ is the disk density of red giants in the solar 
neigbourhood, R is the distance from the Galactic Center within the plane, 
R$_0$~=~8.5~kpc is the distance from the Sun to the Galactic Center, 
h~=~3.5~kpc and h$_z$~=~325 pc for the giants. For r~$<$~R$_0$, 
R--R$_0$~=~--r; r~$>$~R$_0$, \mbox{R--R$_0$~=~r--2R$_0$}; and at 
$b$~=~2$^\circ$, z/h$_z$~=~r/(9.3kpc). The integrations of Eqs. (A.1-2) 
over r along the line of sight ($\ell$, $b$), from 0 to R$_0$ and 
from R$_0$ to $\infty$, respectively, are straightforward, yielding, 
in the direction $\ell$=0, $b$=2$^\circ$, 
N$_{\rm F}$ $\propto ~20 \rho _{0g}$ for the disk counts with 
r~$<$~R$_0$, and N$_{\rm Bk}$ $\propto ~12 \rho _{0g}$ for the disk 
counts with r~$>$~R$_0$.

For the bulge,
 
\begin{eqnarray*}
	\rho _{gB}(r,z) = 3.6\times\rho _{0g}\, x^{-1.8}\exp{(-x^3)} \hskip 2.2cm (\rm A.3)
\end{eqnarray*}

where 

\begin{eqnarray*}
	x & = & \sqrt{R^2\, +\, (1.6z)^2}/R_1^2\\ 
         & = & \sqrt{R^2_0[\ell ^2\, +\, (1.6b)^2]\, +\, r'^2}/R_1^2 \hskip 2.5cm (\rm A.4)
\end{eqnarray*}

where R$_1$~=~2.0~kpc is the bulge ``radius''. However, for the small values 
of $\ell$ \& $b$ in the considered fields, most of the integral of 
$\rho _{gB}$(r,z) comes from bulge points with x~$<<$~1. Therefore, the 
term $\exp{(-x^3)}$ may be neglected in Eq. (A.3). In the 
direction $\ell$=0, $b$=2$^\circ$, the bulge counts are thus proportional to 

\begin{eqnarray*}
	3.6\times\rho _{0g}\, \int_{-\infty}^{+\infty}{x^{-1.8}dr'} \hskip 4.3cm (\rm A.5)
\end{eqnarray*}

or 

\begin{eqnarray*}
& & 	3.6\times\rho _{0g}\, R_1(R_0/R_1)^{-0.8}[b^2+(\ell /1.6)^2]^{-0.4}\\ 
& &  \times \int_{-\infty}^{+\infty}{(y^2+1)^{-0.9}dy} \hskip 4.2cm (\rm A.6)
\end{eqnarray*}

yielding for the bulge counts, in the direction $\ell$=0, $b$=2$^\circ$, 
N$_{\rm Bu}$ $\propto ~57 \rho _{0g}$. 

With this model, on the line of sight $\ell$=0, $b$=2$^\circ$, one thus 
expects to find about 65\% of bulge sources, 22\% of disk sources with 
r~$<$~R$_0$, and 13\% of disk sources with r~$>$~R$_0$, among the ISOGAL 
detected sources.\\

{\bf Remark.} Similar integrals intervene in the computation of the total 
stellar mass column densities M$_{\rm cd}$~=~$\int_{0}^{+\infty}{\rho _{g}(r,z)~ dr}$, 
yielding, with this model: 

\begin{eqnarray*}
& &	M_{\rm cdBu}(M_\odot kpc^{-2}) = \\
& & 100\times \rho _{M0}(M_\odot kpc^{-3})
\times [b^2+(\ell /1.6)^2]^{-0.4} \hskip 1.0cm (\rm A.7)
\end{eqnarray*}

for the mass column density of the bulge, where $b$ and $\ell$ are expressed 
in degrees. $\rho _{M0}$ is the stellar mass density in the solar 
neighbourhood expressed in M$_{\rm \odot}$ per kpc$^3$. We will adopt the 
value of Chabrier (2001) $\rho _{M0}$~=~4.5~10$^7$M$_{\rm \odot} kpc^{-3}$. 
Since most of the bulge sources are close to the Galactic Center, one can 
directly infer the projected bulge mass per solid angle unit, e.g. per square 
degree. For instance, in the direction $\ell$=0, $b$=2$^\circ$, 

\begin{eqnarray*}
	\Sigma _{\rm Bu(0,2)} \approx  5.7\times 10^7 M_{\rm \odot} deg^{-2} \hskip 3.3cm (\rm A.8)
\end{eqnarray*}

\acknowledgements

We thank K.S. Jeong and M. Groenewegen for providing detailed results of 
modelling the infrared emission of AGB stars, and A. Robin \& J. Blommaert for 
helpful discussions.

This research is supported by the Project 1910-1 of Indo-French Center for 
the  Promotion of Advanced Research (CEFIPRA). 

The DENIS project is supported, in France by the Institut National des
Sciences de l'Univers, the Education Ministry and the Centre National de la
Recherche Scientifique, in Germany by the State of Baden-W\"urtemberg, in
Spain by the DGICYT, in Italy by the Consiglio Nazionale delle Ricerche, in
Austria by the Fonds zur F\"orderung der wissenschaftlichen Forschung und
Bundesministerium f\"ur Wissenschaft und Forschung.

This publication makes use of data products from the Two Micron All Sky
Survey, which is a joint project of the University of Massachusetts and the
Infrared Processing and Analysis Center/California Institute of Technology,
funded by the National Aeronautics and Space Administration and the National
Science Foundation.

This research made use of data products from the Midcourse Space Experiment,
the processing of which was funded by the Ballistic Missile Defence 
Organization with additional support from NASA office of Space Science. 

MS is supported by the Fonds zur F\"orderung der wissenschaftlichen
Forschung (FWF), Austria, under the project number J1971-PHY.

\end{document}